\documentclass[12pt,preprint]{aastex}
\newcommand{\velunitns}{km\,s$^{-1}$}
\newcommand{\velunits}{km\,s$^{-1}$ }
\newcommand{\accunits}{km\,s$^{-1}$\,yr$^{-1}$ }
\newcommand{\accunitns}{km\,s$^{-1}$\,yr$^{-1}$}

\newcommand{\about}{$\sim$}

\newcommand{\degrs}{\hbox{$^{\circ}$} }
\newcommand{\degrns}{\hbox{$^{\circ}$}}

\slugcomment{}

\shorttitle{}
\shortauthors{Humphreys et al.}

\usepackage[]{natbib}
\usepackage{color}

\begin{document}


\title{Toward a New Distance to the Active Galaxy NGC~4258:\\ 
II. Centripetal Accelerations and Investigation of Spiral Structure}


\author{E. M. L. Humphreys, 
{M. J. Reid}, 
{L. J. Greenhill},
{J. M. Moran}, \&  
{A. L. Argon}}
\affil{Harvard-Smithsonian Center for Astrophysics, 60 Garden Street,
   Cambridge, MA 02138}

\begin{abstract}
We report measurements of centripetal accelerations of maser spectral components of NGC~4258
for  51 epochs spanning 1994 to 2004.
This is the second paper of a series, in which the goal is determination of a new geometric 
maser distance to NGC~4258 accurate to possibly \about3\%. 
We measure accelerations using a formal analysis method that involves simultaneous  
decomposition of maser spectra for all epochs into multiple, Gaussian components.  
Components are coupled between epochs by linear drifts (accelerations) from their  
centroid velocities at a reference epoch. For high-velocity emission, accelerations lie in the range 
$-$0.7 to $+$0.7 \accunits indicating an origin within 13\degrs of the disk midline (the perpendicular 
to the line-of-sight to the black hole). Comparison of high-velocity emission projected positions in VLBI images, 
with those derived from acceleration data, provides evidence that masers trace real gas dynamics. 
High-velocity emission accelerations do not support a model of trailing shocks associated with spiral arms in the disk. 
 However, we find strengthened evidence for spatial periodicity in 
high-velocity emission, of wavelength 0.75~mas. This supports suggestions of spiral structure due to 
density waves in the nuclear 
accretion disk of an active galaxy. Accelerations of low-velocity (systemic) emission lie in the range  
7.7 to 8.9 \accunitns, consistent with emission originating from a concavity where the thin, warped disk 
is tangent to the line-of-sight. 
A trend in accelerations of low-velocity emission, as a function of Doppler velocity, 
may be associated with disk geometry and orientation, or with the presence of spiral structure.
%%%%%%%%%%%%%%%%%%%%%%%44 

\end{abstract}



%% subject header.  Objects should be in the appropriate "individual"
%% headers (e.g. quasars: individual, stars: individual, etc.) with the
%% additional provision that the total number of headers, including each
%% individual object, not exceed six.  The \objectname{} macro, and its
%% alias \object{}, is used to mark each object.  The macro takes the object
%% name as its primary argument.  This name will appear in the paper
%% and serve as the link's anchor in the electronic edition if the name
%% is recognized by the data centers.  The macro also takes an optional
%% argument in parentheses in cases where the data center identification
%% differs from what is to be printed in the paper.



\section{Introduction}

The Hubble constant (H$_\circ$) is a cornerstone of the extragalactic distance scale (EDS) 
and is a fundamental parameter of any cosmology.  Measured period-luminosity (P-L) relations 
for Cepheid variable stars have been used to determine the EDS, based on estimation of the 
distances to galaxies within $\sim$30~Mpc and calibration 
of distance indicators that are also found well into the Hubble Flow (e.g., the Tully-Fisher relation 
and type-1a supernovae).  
The best present estimate of H$_\circ$ \citep{Freedman2001} is accurate to 10\%, 
72$\pm3$(random)$\pm7$(systematic) km\,s$^{-1}$\,Mpc$^{-1}$.  Other estimates of H$_\circ$ 
are generally consistent, though sometimes model dependent
\citep[e.g., microwave background fluctuations;][]{Spergel2006}
 or somewhat outside the $\pm$10\% uncertainties
 \citep[e.g., H$_\circ$ = 62 km\,s$^{-1}$\,Mpc$^{-1}$;][and references therein]{Sandage2006}.

Several sources of systematic error affect the accuracy of Cepheid luminosity calibrations.  
These include uncertainty in the distance to the Large Magellanic Cloud (LMC), which determines 
the zero point of the P-L relation and for which recent estimates differ by up to $\pm$0.25 mag, or $\pm$12\% 
\citep{Benedict2002}, and uncertainty in the impact of metallicity on the P-L relation 
\citep[e.g.,][]{Udalski2001,Caputo2002,Jensen2003}.
This is particularly important because the LMC is metal-poor with respect to other 
galaxies in the Freedman et al. study.  For a description of controversy concerning the estimation of 
H$_\circ$ see \citet{Macri2006} and \citet[][hereafter Paper I]{Argon2007}.  

For the nearby active galaxy NGC~4258, independent distances may be obtained from analysis of Cepheid brightness 
and water maser positions, velocities, and accelerations.  Comparison of the distances would enable refinement of 
Cepheid calibrations and the supplementation or replacement of the LMC as anchor, thus reducing uncertainty in H$_\circ$ 
and the EDS.  Implications may include a better constraint of cosmological parameters, such as the flatness 
of the Universe and the 
equation of state for dark energy \citep[e.g.,][]{Hu2005,Spergel2006}, which would discriminate among different 
origins, e.g., the cosmological constant and quintessence.

Compact maser emission in NGC~4258, from the 6$_{16}$-5$_{23}$ transition of water (22235.080 MHz) was 
first detected near to the systemic velocity (v$_{sys}$) by \citet{Claussen1984}, where 
v$_{sys}$ = 472$\pm$4 \velunits 
\citep[referenced to the Local Standard of Rest (LSR), and using the radio 
definition of Doppler shift;][]{Cecil1992}. The discovery of high-velocity emission at \about v$_{sys}\pm$1000 \velunits provided critical evidence that maser emission probably arises from material in orbit about a 
massive central black hole \citep{Nakai1993}. Early Very Long Baseline Interferometry (VLBI) observations 
showed that masers delineated an almost edge-on, sub-parsec-scale rotating disk \citep{Greenhill1995a}. 
Studies of maser centripetal accelerations, inferred from secular velocity drifts in the peaks of 
spectral components, produced independent evidence of a disk geometry 
\citep{Haschick1994,Greenhill1995b,Nakai1995,Watson1994}. Low-velocity masers 
(near to v$_{sys}$) displayed positive drifts of $\sim$9 \accunitns, which located them 
on the front side of the disk, whereas high-velocity masers drifted by $<$ $\pm$1 \accunitns, 
confining emission to lie close to the disk midline (the diameter perpendicular to the line of sight). 
Very Long Baseline Array \footnotemark[1] (VLBA) studies provided conclusive evidence supporting the disk model, 
established a Keplerian rotation curve to better than 1\% accuracy for maser emission, 
and traced a warp in the disk structure (\citealt{Miyoshi1995}; \citealt{Herrnstein1996,Herrnstein2005}; 
 and early reviews of \citealt{Moran1995,Moran1999}). 

\footnotetext[1]{The Very Long Baseline Array and Very Large Array are operated by the National Radio Astronomy Observatory, a facility of the National Science Foundation operated under cooperative agreement by Associated Universities, Inc.}

A geometric distance can be obtained by modeling the 3-D geometry and dynamics traced 
by the maser line-of-sight (LOS) velocities, positions, and LOS accelerations or proper motions.  
Assuming circular orbits, \citet{Herrnstein1999} obtained the most accurate distance to 
NGC~4258 thus far, $7.2\pm0.2$(random)$\pm0.5$(systematic)~Mpc.  The total error is 7\%, and the 
systematic component largely reflects an upper limit on the eccentricity of 0.1 for confocal 
particle orbits.  This contributes 0.4 Mpc of the systematic uncertainty budget.

This paper is the second of a series in which we revisit estimation of the maser 
distance to NGC~4258, using an expanded dataset and more detailed analyses.  Our goal
is to reduce the uncertainty by a factor of 2-3, through reduction in both random and systematic errors.  \citet{Herrnstein1999} used four epochs of VLBA data spanning 3 years to estimate distance.  To this, we have added 18 new VLBI epochs over 3.4 years (Paper I).  These combined data will facilitate a more thorough disk modeling, including the incorporation of orbital eccentricity and periapsis angle as parameters.  Here, we address the measurement of radial accelerations for maser emission from a time-series 
of spectroscopic measurements, combining VLBI, Very Large Array\footnotemark[1] (VLA), and single-dish spectra.  Future work will present the model of the cumulative data (position, velocity, and acceleration) to obtain a best-fit distance.  

In a parallel series of papers starting with \citet{Macri2006}, analysis of recent 3-color, 
wide-field Hubble Space Telescope (HST) Cepheid photometry is used to estimate an improved 
``standard candle'' distance to NGC~4258 \citep[cf.][]{Newman2001}, with supplementary information 
from near-infrared and high angular resolution data that place tighter constraints on systematics (crowding and extinction).  

Our approach to the measurement of accelerations is very different from that of previous work, 
wherein individual maser spectral Doppler components were identified by local maxima in blended 
line profiles at each epoch.  Components were matched up among different epochs on a maximum likelihood 
basis or in ``by eye'' analyses, depending on the study, in order to determine drifts in component peak or 
fitted centroid velocities \citep{Haschick1994,Greenhill1995b,Nakai1995,Herrnstein1999,Bragg2000}. 
For low-velocity emission 
in particular, this method is subject to ambiguity in component identification unless the spacing between epochs 
is short ($<3$-4 months), because of intensity fluctuations among blended, drifting components.  This informal 
fitting technique is also susceptible to biases in the velocities of local maxima due to blending.   Here, we 
decompose the spectra into individual Gaussian components for many epochs simultaneously, constrained by a separate 
linear velocity drift in time for each feature.  The constraint greatly increases the robustness of the decompositions, 
 is less subjective than ``by eye'' techniques, and enables the discernment of weak spectral features with greater confidence. Including the VLBI monitor data reported in Paper I, we applied this fitting to 36 spectra of low-velocity emission spanning $\approx $6 years, and 40 spectra of high-velocity emission spanning $\approx $10 years.

Implicit in the estimation of distance is construction of a disk dynamical model from acceleration, 
position, and velocity measurements. These may also be used to detect disk sub-structure.  
Analysis of structure on the smallest scales enables identification of maser spots as representative 
of physical clumps. Clustering of maser spots enables tests of predicted spiral structure induced by 
gravitational instabilities.  \citet{Maoz1995} proposed that spiral density waves form in the NGC~4258 disk, 
and the high-velocity masers mark the density maxima along the disk midline, which is where gain paths are greatest.  \citet{Maoz1998} proposed that the spiral pattern 
is traced by shocks, with maser emission arising in relatively narrow post-shock regions and visible only 
when the spiral arms are tangent to the line-of-sight.  Due to the pitch angle of the spiral pattern, the masers 
would arise systematically offset from the midline, and the acceleration signatures predicted by the two models 
enable discrimination \citep{Bragg2000}.  

We describe the spectroscopic observations with which accelerations are estimated in Section~\ref{s:data}.  
Decomposition of spectra into Gaussian components whose velocities drift in time and the inferred 
accelerations are described in Sections~\ref{s:spectrumdecomp} and~\ref{s:results}.  Comparisons with 
previous work are in Section~\ref{s:sect5}.  
Implications of the new analyses for disk structure follow, with focus on physical identification of 
maser clumps, signatures of spiral structure, and strong constraints on orbital eccentricity.
The next paper in this series will give the improved distance to NGC~4258 estimated
from analyses of maser data presented in Paper I and this paper.
  
\section{The Dataset}
\label{s:data}

\begin{figure}
\epsscale{0.5}
%\plotone{fig1_jun19.vps}
\plotone{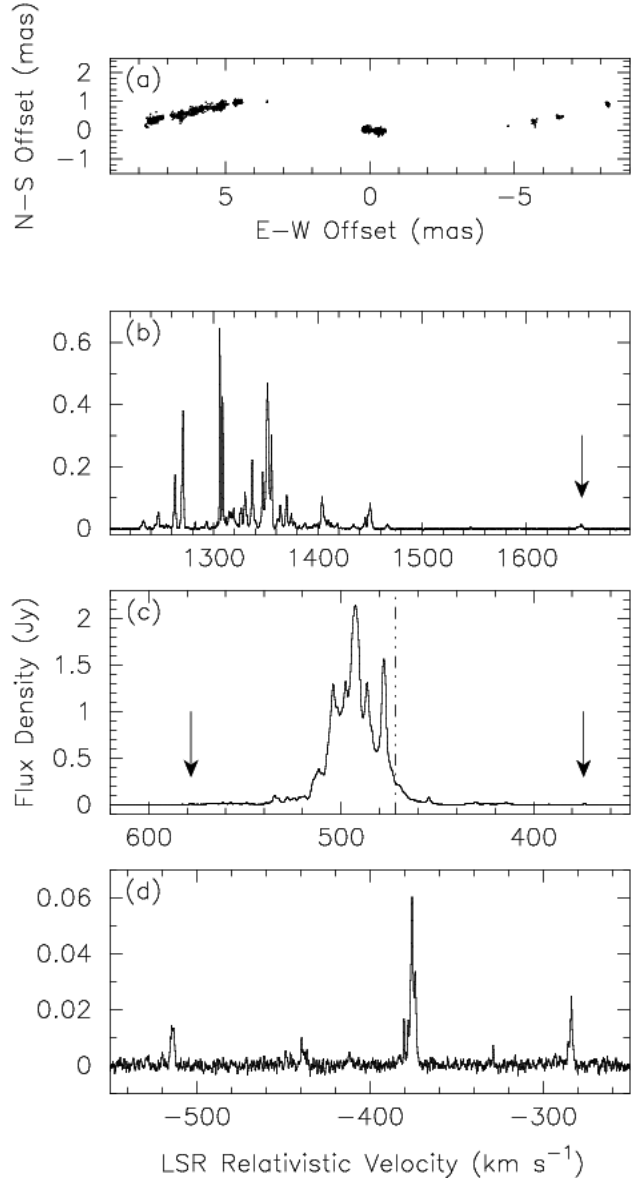}
\caption{\footnotesize {\it (a)}
Sky positions for 22~GHz maser emission in NGC~4258 obtained using VLBI on 
18 epochs from 1997 March 06 to 2000 August 12 described in Paper I. Red-shifted 
high-velocity emission occurs from 3 to 8 mas in the disk; low-velocity emission 
occurs in the range -1 to 1 mas and blue-shifted high-velocity emission in the range 
5 to 9 mas. Note that $(0,0)$ in the plot does not represent the disk dynamical center. 
{\it (b)} GBT spectrum of red-shifted high-velocity emission from \citet{Modjaz2005}, 
obtained on 2003 October 23. The 1$\sigma$ noise in the spectrum is 1.6 mJy. The arrow marks 
weak, newly-detected emission at 1652 \velunits (Paper I).  
{\it (c)} GBT spectrum of low-velocity 
maser emission. The arrows mark emission extrema. The dotted line denotes the galactic 
systemic velocity of 472 \velunits \citep{Cecil1992}. 
{\it (d)} GBT spectrum of blue-shifted maser emission.
\label{f:figure1}}   
\end{figure}

\begin{figure}
\epsscale{0.7}
%\plotone{systemicsandreds.ps}
\plotone{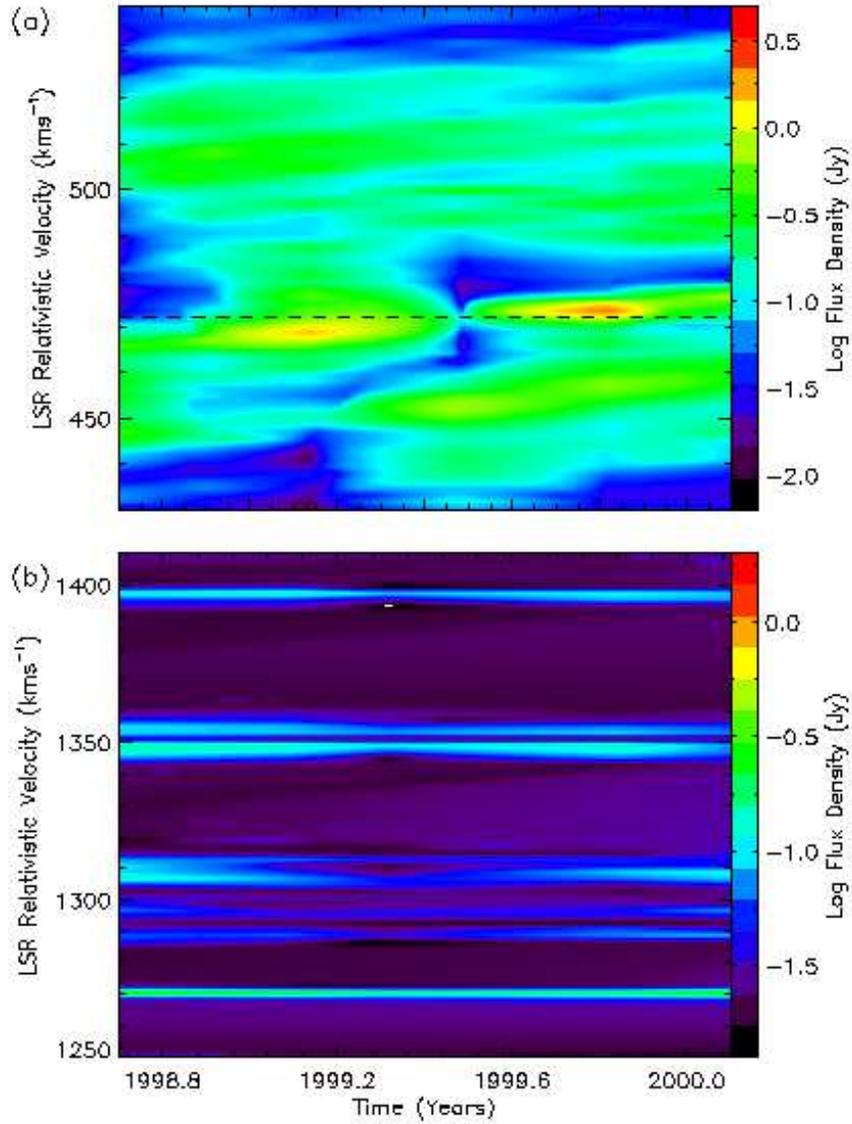}
\hspace{-1cm}
\caption{\footnotesize {\it (a)} 
Velocity vs. time plot for systemic components using 
data from 1998 January 27 to 2000 August 12 during which sampling 
was every \about3 months. Drifts in Doppler velocities of strong components 
are clearly evident and consistent with centripetal accelerations. The dashed 
line marks the galactic systemic velocity of 472 \velunitns.
{\it (b)} Corresponding plot for red-shifted components. Small magnitudes of 
drifts in component Doppler velocities are consistent with near-zero observed 
accelerations for components near the disk midline. 
\label{f:figure2}}
\end{figure}

The primary purpose of this paper is to report the measurements of maser component accelerations with high accuracy 
to be used in determination of a new geometric distance to NGC~4258. We attempted to 
reduce statistical uncertainties in the accelerations by inclusion of a larger number of 
epochs over a longer time baseline than that done by \citet{Herrnstein1999}. In this study, we analyzed 
spectra of water maser emission at 22.235 GHz from NGC~4258 at fifty-one epochs, between 1994 April 19 
to 2004 May 21. These data were presented in detail in Paper I. We summarize the observations in 
Table~\ref{t:dataset} and display subsets 
of the data in Figures~\ref{f:figure1} and~\ref{f:figure2}.

A compilation of the spectra measured from 1994 April 19 to 1997 February 10 was 
presented by \citet{Bragg2000}. Observations were made approximately 1 to 2 months using 
the VLBA (5 epochs), the VLA (17 epochs) and the Effelsberg 100-m telescope of the 
Max-Planck-Institut f\"{u}r Radio Astronomie (5 epochs). The 1$\sigma$ noise in the 
spectra range from 15 to 110 mJy for channel spacings of $<$0.35 \velunits (Table~\ref{t:dataset}). 
At the majority of epochs using the VLA and VLBA, both low- and high-velocity emission 
were measured, however some exceptions are noted in Table~\ref{t:dataset}. Only the red-shifted high-velocity 
spectrum was observed at Effelsberg.

The spectra from eighteen VLBI epochs, measured between 1997 March 06 to 2000 August 12, 
are displayed in Paper I. Twelve of the epochs used the VLBA only (the ``medium-sensitivity'' epochs) 
and had 1$\sigma$ noise in the range 3.6 to 5.8 mJy for 1.3 \velunits wide channels.  At 
medium-sensitivity epochs, low-velocity emission plus either red-shifted or blue-shifted 
high-velocity emission were observed. The observing set-up was designed to enable exploration 
of a continuous range of velocities between known spectral-line complexes, and a limited range to higher velocities.  
The remaining six epochs 
(the ``high-sensitivity'' epochs) involved the VLBA, the phased VLA and Effelsberg. 
The instantaneous bandwidth of the high-sensitivity observations was sufficient to perform 
simultaneous imaging of known low- and high-velocity emission, and resulted in 1$\sigma$ noise 
of  2.3 to 4.7 mJy per \velunitns. 
The average time between the observations was \about2.5 months. We extracted spectra from 
VLBI images by fitting 2-D Gaussian brightness distribution functions to all peaks in velocity channel maps. 
For a detailed description of the data analysis method, see Paper I. 
 Of the remaining epochs, three were obtained using the Green Bank Telescope 
(GBT) between 2003 April 10 and 2003 December 8, with 1$\sigma$ noise levels 
of \about3 mJy per 0.16 \velunits channel (Modjaz et al.\,2005; Modjaz, priv.\,comm.\,2005; Kondratko, priv.\,comm.\,2006). 
We obtained the final epoch of data on 2004 May 21 
using the VLA, for limited portions of the red-shifted high-velocity spectrum only, and with 
an r.m.s noise of 20 mJy per 0.21 \velunits channel.

\section{Multi-Epoch Spectrum Decomposition}
\label{s:spectrumdecomp}

To measure accelerations of maser components, as well as amplitude and line width variations during
the monitoring, 
we decomposed portions of spectra from multiple epochs simultaneously into individual Gaussian line profiles. 
We identified a component at different epochs by solving 
for its constant drift from a centroid velocity 
at a reference time. 
We could therefore account for each Gaussian at every epoch, 
even if the amplitude 
had fallen below our detection threshold, which minimized potential 
for confusion among components.

We employed a non-linear, multiple Gaussian-component least-squares $\chi^2$ minimization 
routine that we dimensioned for a maximum of 84 time-varying Gaussians and 
40 epochs of data in any given fit, which was limited by computational factors. Each maser component was represented by 
a Doppler velocity $v_{los}(t_{0})$ at a reference time $t_{0}$, a linear velocity drift $\dot{v}_{los}$, 
and a line width and amplitude at every epoch. We could choose to solve 
for constant, or time-varying, component line widths. We set {\it a priori}  
line widths in order to prevent large deviations from physical values 
expected for maser line widths at gas kinetic temperatures of 400 to 1000~K. 
The {\it a prioris} were included as extra 
data points in the fit, and typically prevented deviations of greater than four times the {\it a priori} uncertainties. 
The reference time was chosen to be near the middle of the monitoring period. 
All velocities in this paper are quoted for a $t_{0}$ of 1999 October 10, or monitoring day 2000, unless stated otherwise.

We note that fitting of linear velocity drifts is an approximation for particles on circular orbits 
in a disk in which $v_{los}=v_{rot}\cos(\Omega \Delta t+\phi_0) + v_{sys}$, where v$_{rot}$ 
is the rotational velocity at any given $r$, $\Omega$ is the angular velocity, 
$\Delta t=t - t_{0}$ and $\phi_0$ is the angle from the midline at a reference time t$_{0}$. 
From Monte Carlo simulations, we estimate that the maximum error 
we introduce by using this approximation is 0.1 \accunitns. We chose the linear approximation in order to determine 
velocity drifts without any model assumptions. 
 
\subsection{High-Velocity Emission}

\begin{figure}
\epsscale{1.0}
\includegraphics[angle=-90,scale=0.8]{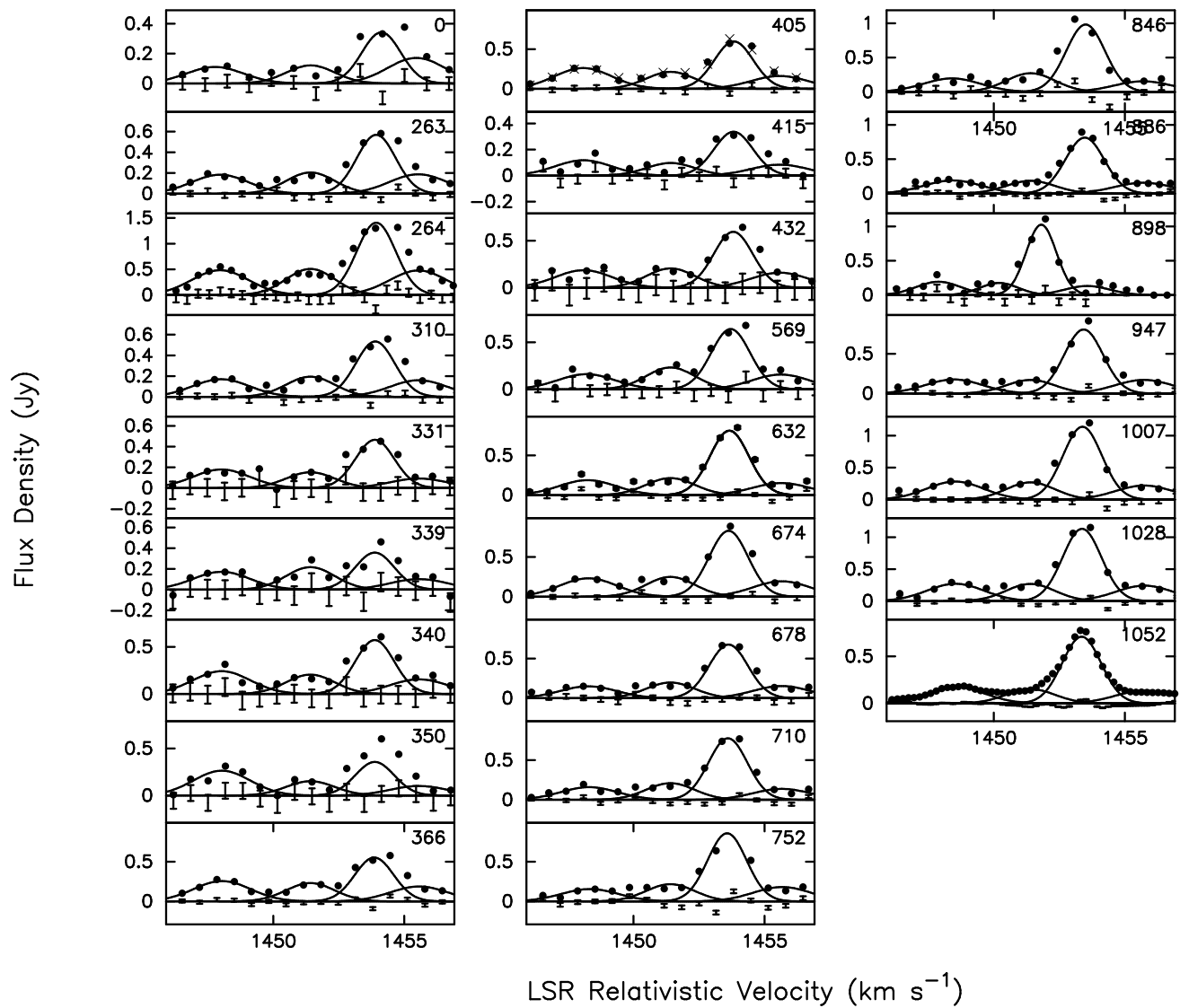}
\caption{Example of a simultaneous 4-Gaussian function fit to a blend
of red-shifted components
over 25 epochs (2.9 years). Filled circles mark the data points. Residuals are
plotted with 1$\sigma$ errorbars.
Model Gaussian functions are shown in solid black lines. The epoch day 
number is marked in the upper right of each panel. Table~\ref{t:dataset}
 lists the corresponding dates.
\label{f:redfit}}
\end{figure}

\begin{figure}
\epsscale{1.0}
%\plotone{feature-280_bw2.vps}
\plotone{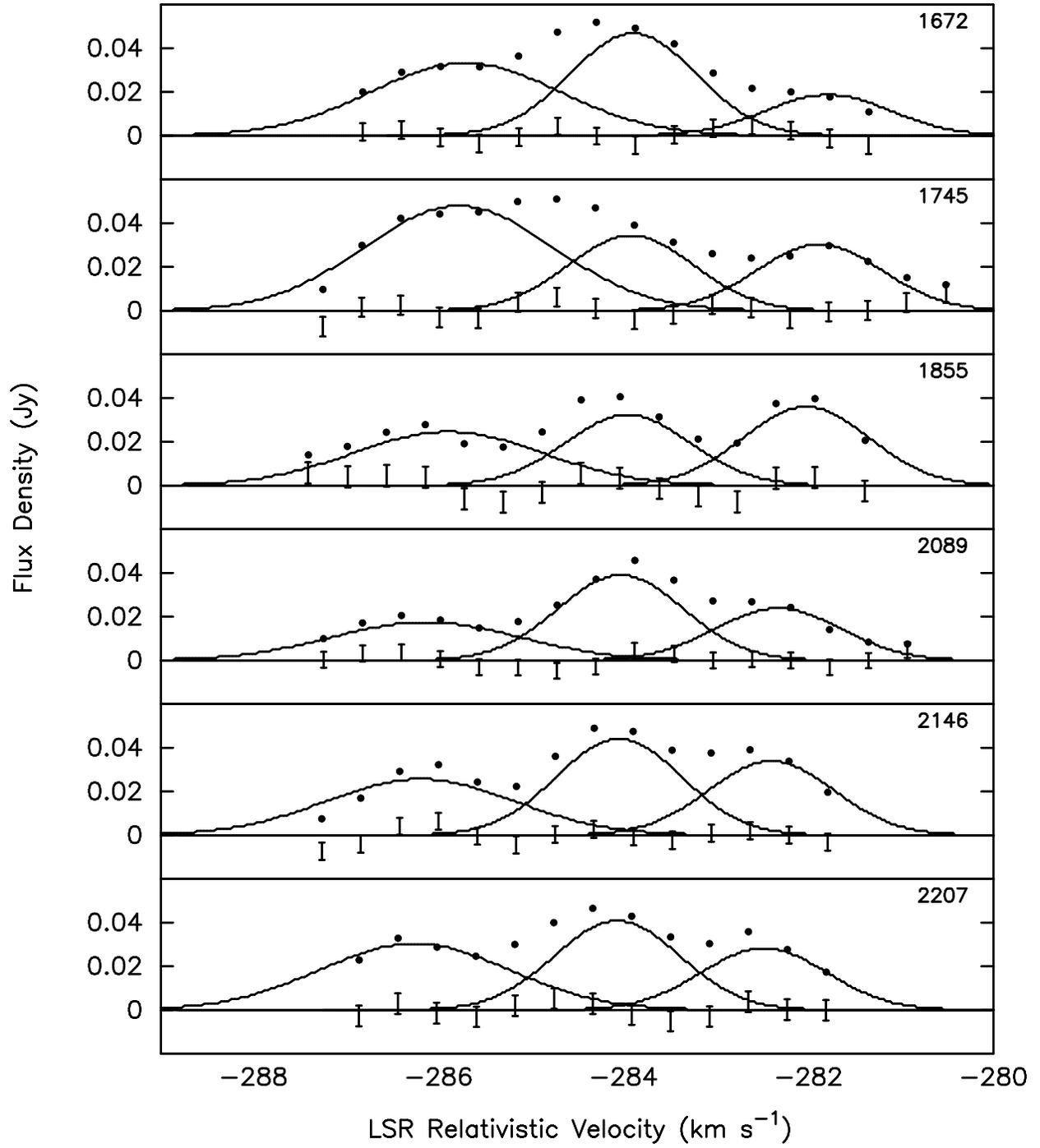}
\caption{As for Figure~\ref{f:redfit}, but for a blue-shifted high-velocity emission
blend. 
\label{f:bluefit}}
\end{figure}

The high-velocity spectrum consists of isolated blends of small numbers of components, 
with low drift rates (Figures~\ref{f:figure1} and~\ref{f:figure2}).  To facilitate the decomposition, we divided 
the high-velocity spectra into individual blends, of typical velocity extent 10 to 20 \velunitns. We performed the fitting of each blend 
iteratively. First, we identified the number of prominent peaks in the blend at a 
high-sensitivity and high-spectral resolution epoch (e.g., 1998 September 5). 
We fit this number of Gaussian functions to the data over all epochs and examined the residuals at each epoch. Where 
deviations in the residuals exceeded 5$\sigma$, we refit the segment using more Gaussian functions. 
We repeated this procedure until there was no systematic structure remaining in the residuals 
above the 5$\sigma$ level. We solved for constant line widths as a function of time for high-velocity 
emission, because the signal was typically $< $1~Jy,  and line widths at any given epoch were 
otherwise not well-constrained. The fits for each high-velocity blend included between four and 
forty epochs of data, covering time baselines of 0.5 to 9 years (Tables~\ref{t:table2} and~\ref{t:bluetable}) depending on 
component lifetime and time-sampling. Example fits are shown in Figures~\ref{f:redfit} and~\ref{f:bluefit} for red- and 
blue-shifted emission, respectively.

\subsection{Low-Velocity Emission}

The low-velocity emission spectrum consists of strong (typically 2 to 10~Jy), highly-blended emission 
for which previous work has shown components drift at a mean rate of 
\about9 \accunits \citep{Haschick1994,Greenhill1995b,Nakai1995,Herrnstein1999,Bragg2000}. 
Some studies
note a systematic trend in accelerations of low-velocity emission
as a function of Doppler velocity, with accelerations of 8.6$\pm$0.5 \accunits at velocities less than 470 \velunitns, and
 of 10.3$\pm$0.6 \accunits at velocities greater than 470 \velunits \citep{Haschick1994,Greenhill1995b}. Others 
do not draw attention to such an effect \citep{Nakai1995,Herrnstein1999,Bragg2000}.   
We therefore approached decomposition of the low-velocity spectra using two methods: (i) assuming
that the linear velocity drift approximation holds over the entire 36 epoch monitoring, and including all 36 epochs
of data; and, 
(ii) assuming that there may be changes in accelerations during the monitoring period, 
and dividing the data in four consecutive datasets of nine epochs, for which fitting was performed
independently. When weighting the data for fitting, we added 2\% of the channel flux density 
in quadrature to the 1$\sigma$ noise, to account for the dynamic range limitations of our observations (see Paper I).

To perform the 36-epoch fitting, we first decomposed the 
spectrum into a {\it preliminary} set of Gaussian components at one epoch only. We selected 
a high-spectral resolution and high-sensitivity epoch for this purpose (1998 September 5). 
We performed a fit over a subset 
of 6 adjacent epochs for the low-velocity spectrum as a whole, to obtain preliminary values of acceleration. 
We increased the number of Gaussian functions used in the fitting with each successive iteration, adding additional Gaussians to the fit, until no 5$\sigma$ systematic deviations 
remained (in the residuals).
This analysis provided a set of ``seed'' parameters 
to perform a new decomposition over 36 epochs. For ease of computation, 
we divided the spectrum into segments of \about12 Gaussian components 
and performed the decomposition of each 
segment iteratively. We used a moving time window 
to select the relevant data range for segments at different 
epochs in order to track the $\sim$9 \accunits accelerations. At segment edges, we overlapped 
fitted regions in each case 
by 2 to 3 components to ensure that fits were 
consistent across the low-velocity spectrum. 
Using this method, we obtained residual deviations of $<$ 5$\sigma$
everywhere except at the time extrema of our dataset, where we obtained significantly larger deviations
in residuals.

For the 9-epoch fits, we also decomposed the 
spectrum into a preliminary set of Gaussian components at one epoch only (chosen to be near the middle of each dataset). 
We then performed an initial fit over the entire 9 epochs to obtain component accelerations.
Unlike the 36-epoch fitting, it was not necessary to sub-divide
the spectrum and use a moving time window, and we performed the decomposition
across the low-velocity spectrum as a whole in each case.
We increased the number of Gaussians in each of the fits iteratively, examining 
the residuals after each fit, and adding components where large deviations occurred.
We repeated this process until no $>$ 5$\sigma$ systematic deviations remained in the residuals. 
In each of the 9-epoch fits, this required \about55 Gaussians of line width 1 to 4 \velunitns.
The four, independent fits to the data indicated that there is a persistent trend of component 
accelerations as a function of Doppler velocity (or equivalently time), explaining why a good fit was not obtained in the
longer time-baseline, 36-epoch method. We adopt results from the 9-epoch fits in the sections
that follow.

\section{Results}
\label{s:results}

\subsection{High-Velocity Emission}
\label{ss:sect4.1}

We measured accelerations for 24 red- and 8 blue-shifted high-velocity components 
and found them to range between $-$0.40 and $+$0.73 \accunitns, and between 
$-$0.72 and $+$0.04 \accunitns, respectively (Tables~\ref{t:table2} and~\ref{t:bluetable}). 
We found a weighted 
average of red-shifted high-velocity accelerations of 0.02$\pm$0.06  \accunitns, 
and a weighted average blue-shifted high-velocity accelerations 
of $-$0.21$\pm$0.08 \accunitns. We note that there is a non-normal distribution 
with respect to acceleration systematics due presumably to disk structure. We estimated the azimuth angle, $\phi$, 
of maser components from the disk midline for a flat disk model in terms of measurable quantities; 
the line-of-sight velocity $v_{los}$ with respect to v$_{sys}$ and line-of-sight acceleration $\dot{v}_{los}$, 
adopting $\phi \approx \sin \phi / \cos^4 \phi = GM_{BH} \dot{v}_{los} / v^4_{los}$ 
\citep{Bragg2000}, where $G$ is the gravitational constant, $M_{BH}$ is the 
central mass and where $\phi$ is measured from the midline for red-shifted 
high-velocity emission, increasing in the sense of disk rotation. For red-shifted 
high-velocity emission, we found that azimuth angles lie in the range -8.9\degrs to 
+12.6\degrs and have a mean and standard deviation of 0.2$\pm$3.6$\rm ^o$, i.e., 
are centered on the midline. For blue-shifted high-velocity 
components we determined that azimuth angles lie in the range 166.8\degrs to +180.5\degrs 
and have a mean value of 176.7$\pm$6.9$\rm ^o$. Since for a warped disk $\phi$ $\propto$ $\sin^3 i$, there
is a 0.2 to 2\% error in these values \citep{Herrnstein2005}. 
Using a flat, edge-on 
Keplerian disk model and distance of 7.2~Mpc \citep{Herrnstein1999}, we derived 
disk radii of high-velocity emission in the range 0.17 to 0.29~pc. 
We measured line widths in the range 1.0 to 5.0 \velunits for the high-velocity emission. 

In Paper I, Argon et al.\,(2007) reported the discovery of new red-shifted 
high-velocity emission at 1562 \velunitns, and also detection of emission at 1652 \velunits previously
discovered at the GBT by \citet{Modjaz2005}.  We were 
unable to estimate the accelerations of these components due to blending and the limited data available. 

\subsection{Low-Velocity Emission} 
\label{ss:sect4.2}

We measured accelerations for four, time-consecutive datasets consisting of 9 epochs each. 
We found a significant, reproducible trend in acceleration as a function of 
component Doppler velocity for each data subset with a systematic variation of  1 \accunits across
the low velocity emission range of $\sim$430 to 550 \velunitns.
We binned the data from all four fits to yield measured accelerations 
in the range 7.7 to 8.9 \accunits for 55 low-velocity components (Figure~\ref{f:systrend}; Table~\ref{t:systable}).
 The acceleration trend may be associated with a systematic change in the radius
at which maximum velocity coherence is achieved due to the disk warp (see Section~\ref{s:implications_lv}). 
We performed a weighted fit to accelerations as a function of velocity and computed a
PV diagram that would be consistent from $v_{los}=\sqrt{GM/r^3}b$ where $b$ is the
impact parameter and $a_{los}=(GM/r^2)\sin\phi$ for $\phi$=270\degrs and assuming circular rotation. The predicted
and observed PV diagrams agree to within 1$\sigma$ for velocities from 390 to 540 \velunitns (Figure~\ref{f:systrendpv}).
For the disk shape of \citet{Herrnstein2005}, 
we find the maser components
lie in a range of deprojected disk radii of 3.9 to 4.2~mas as shown in Figure~\ref{f:velcontours}
(0.14 to 0.15~pc at a distance of 7.2 Mpc)
 i.e., $<$10\% of the radial extent of observed
high-velocity emission.
We discuss possible physical origins for the trend in Section~\ref{s:implications_lv}.

\section{Comparison with Previous Work}
\label{s:sect5}

Over a monitoring period of ten years, 
a high-velocity component orbiting with a period of 700 years travels 5\degrs 
in disk azimuth angle, almost entirely along the line-of-sight. In theory, maser components 
could therefore cross the midline during our monitoring, or during the time gaps among studies, 
and undergo a change in the sign of acceleration. 
However, a comparison of the different acceleration studies (Table~5) shows that
they are in broad agreement. In particular, we compare our acceleration measurements for 
high-velocity emission with those of \citet{Bragg2000} and \citet{Yamauchi2005} (Figure~\ref{f:acccomp}). 
The main difference between our results and those of previous studies is that we measure accelerations 
for more components due to the greater sensitivity and spectral resolution of our data, and due to the 
greater accuracy of the technique we employ to obtain more complete spectrum decomposition. 

High-sensitivity and resolution are of particular importance with respect to blue-shifted high-velocity components. 
 The addition of the VLBI data of Paper I enabled measurement of accelerations for 
8 blue-shifted components, whereas \citet{Bragg2000} measured accelerations for 
 2 blue-shifted high-velocity components only: 0.043$\pm$0.036 \accunits at -440 \velunits 
and -0.406$\pm$0.070 \accunits at -434 \velunitns. \citet{Yamauchi2005} 
obtained a single measurement of 0.2$\pm$0.1 \accunits for emission at -287 \velunitns. 
We obtain similar accelerations for the components also measured by \citet{Bragg2000}. 
However, we spectrally resolve the -287 \velunits emission into 3 components, with 
accelerations of  -0.47, -0.12 and -0.38 \accunits (Figure~\ref{f:bluefit}). 
Fluctuations in blended features may have led 
\citet{Yamauchi2005} to obtain a small positive apparent drift rate. If we fit the 
emission at -287 \velunits as a single Gaussian component, we also obtain a 
positive acceleration in agreement with \citet{Yamauchi2005} within the uncertainties.

The centripetal accelerations for low-velocity emission lie in the same range as 
those measured in previous studies \citep{Haschick1994,Greenhill1995b,Nakai1995,Herrnstein1999,Bragg2000}, 
although
binned measurements of the present work tend to be lower at any given velocity, see Figure~\ref{f:lowvelacccomp}.
This could be due to components at larger radii in the disk, or could be associated with the more robust
measurement technique employed here.  
A trend in accelerations as a function of component Doppler velocity, with the same sign and 
similar magnitude as that measured here, is also reported by \citet{Haschick1994} and by 
\citet{Greenhill1995b}, and is evident in the acceleration data of \citet{Nakai1995} and
\citet{Herrnstein1999}.
 The geometric model of the maser disk by \citet{Herrnstein2005} 
indicates that it is warped, both in inclination and position angles, 
the result of which is to place low-velocity masers in a concavity on the 
front side of the disk (Figures~\ref{f:herrndisk} and~\ref{f:velcontours}). The range of accelerations for 
low-velocity emission determined by this, and other 
studies, could imply a radial spread in low-velocity masers
over the relatively broad bottom of the concavity, which is 
consistent with the model \citep{Herrnstein2005} and the PV diagram (Figure~\ref{f:systrendpv}).

\section{Implications for the Accretion Disk from High-Velocity Emission}

\subsection{Disk Geometry}

We can use the accelerations to confirm that the maser dynamic observables 
reflect real gas dynamics in the NGC~4258 accretion disk.
The small accelerations of the 
high-velocity components suggest that they lie close to the disk midline. 
For this discussion, it is important to visualize the shape of the disk 
in the vicinity of the midline, which is in plane of the sky that includes the 
black hole. We show the geometry of the maser disk determined by 
\citet{Herrnstein2005} in Figure~\ref{f:herrndisk}. 
On the midline, the disk shape is defined by the position angle warp, 
as shown in Figures~\ref{f:redsky} and~\ref{f:bluesky} respectively for the red- and blue-side of the disk. 
That is, the expected midline curve is $y=r\cos\alpha_{r}$, where $y$ is the projected 
vertical position of the warped disk 
on the sky relative to the disk dynamical center in units of milli-arcseconds, $\alpha_{r}$ is 
the disk position angle measured North of East in the plane of the sky (the x-y plane
in Figure~\ref{f:herrndisk}) given by 
$\alpha_r=65.6$\hbox{$^{\circ}$}$+5.04$\hbox{$^{\circ}$}$r-0.22$\hbox{$^{\circ}$}$r^2$
\citep{Herrnstein2005}, 
and where $r$ is the radial position of masers in the disk in mas. We take $r$ \about $b$, 
where $b$ is the impact parameter in the plane of the sky measured from the disk dynamical 
center.

Over the radial range of emission from 4 to 8~mas on the red-side of the disk, the position 
angle varies from 84\degrs to 92\degrns. The ``tilt'' of the disk at the midline 
is given by the inclination warp, which is described by $i_r = 107$\hbox{$^{\circ}$}$ - 2.29$\hbox{$^{\circ}$}$ r$  
\citep{Herrnstein2005}. Hence, for the red side of the disk 
$i_{r}$ varies from 97\degrs at 4~mas, to 89\degrs at 8~mas. The range is the same 
for the blue-side of the disk. It is important to note that, at radii of 7.4~mas, we view 
the disk exactly edge on. Inside this radius, the near-side of the disk is 
tipped down with respect to the observer, while outside it is tipped up. Masers at radii smaller 
than 7.4~mas, that lie above the midline in VLBI images, should be behind the midline plane in the disk and have negative 
accelerations. 
 Similarly masers 
beyond 7.4~mas radius, that appear above the midline in Figures~\ref{f:redsky} and~\ref{f:bluesky}, 
are expected to be in front of the midline plane and should have positive accelerations. 

We compared the vertical 
deviations in the VLBI image caused by the disk projection on the sky (see Figures~\ref{f:redsky} 
and~\ref{f:bluesky}), $y_{image}$, 
with the vertical deviations expected from 
the accelerations, $y_{acc}$. To derive $y_{acc}$ for each maser, we first computed 
its distance in the $z$-direction from the midline from its azimuth (which comes from its acceleration), that is 
$z = b \tan \phi$ where $\phi$ is the azimuth angle of the maser listed in Tables~\ref{t:redtable}
and~\ref{t:bluetable}. 
From this azimuth offset, and the local slope of the warped disk at radius $r$,  $y_{acc}$ 
was calculated as the projected offset from the midline. The plot of $y_{acc}$ vs.
$y_{image}$ is shown in Figure~\ref{f:jim}. The errors 
in $y_{image}$ are the quadrature sum of the errors in the VLBI position and 
the errors in the midline definition.  The errors in $y_{acc}$ are the result of 
acceleration measurement errors, inclination model errors, and errors due 
to a finite vertical distribution of masers. We adopt 
a disk thickness of 12 $\mu$as, the value determined from
low-velocity emission in Paper I.
The disk 
thickness corresponds to a temperature of about 600 K for hydrostatic equilibrium.
 In order to make the reduced $\chi^2$ of the
fit equal 1, it was necessary to add a systematic r.m.s. acceleration of 0.1 \accunitns.  
Since the magnitude of the maser accelerations is up to \about10 \accunitns, 
the systematic errors in the acceleration measurements are at the 1\% level.
The errors in Figure~\ref{f:jim} reflect 
these contributions. Note that there is one significantly 
deviant point at 2.8$\sigma$ (the one at 1271 \velunits and $r=$7.37~mas).

\subsection{Spiral Structure} 

Accelerations and positions of high-velocity emission can also be used to 
investigate theories of spiral structure in the 
accretion disk of NGC~4258. Firstly, \citet{Maoz1995} noted
 periodic clustering of high-velocity component radii in a visual analysis of
 the single VLBI map by \citet{Miyoshi1995}. Maoz inferred a characteristic wavelength 
of $\lambda_{char}=$0.75~mas (0.027~pc) for the clustering, and attributed it to spiral density waves 
that form by swing amplification.  Spiral density waves can form in a disk that is 
unstable to non-axisymmetric perturbation in the 1$< Q \le$2 regime, 
where $Q$ is the Toomre stability parameter \citep{Toomre1964}. {\it Swing} refers to the 
fact that a non-axisymmetric disturbance is converted to a trailing spiral arm (in which the outer tip points backwards, opposite the direction of disk rotation) due 
to shearing by differential rotation. {\it Amplification} occurs during this process, 
since the direction of the shear flow of the arm is the same direction as local epicyclic 
motions, which enhances the self-gravity of the arms. 
To satisfy this stability regime, \citet{Maoz1995} required
densities in spiral arms of n(H$_{2}$) 
= 10$^{8}$ to 10$^{10}$~cm$^{-3}$ which are sufficient to produce 22~GHz maser emission   \citep[e.g.][]{Cooke1985}, whilst those 
in the inter-arm regions are not.
In the theory of \citet{Maoz1995}, masers are discrete 
physical clumps that can move freely through different spiral arms and can be tracked 
between epochs.

\citet[][hereafter MM98]{Maoz1998} also propose that maser emission 
could arise behind spiral {\it shocks} in the NGC~4258 disk, from density and velocity coherence arguments. The shocks 
occur at the interface of spiral arms sweeping through the disk. In 
this case masers may not be discrete entities but a wave phenomenon, 
with emission arising at the tangents of the spiral shocks along the 
LOS, where the path length for amplification is highest. At 
different epochs, maser emission would not originate 
from the same gas condensations. The accelerations of the masers would no longer 
reflect pure Keplerian dynamics, as the action of the shock causes the excitation 
points for masers to move to greater radii in the disk. All maser velocities  
would shift to lower absolute values. This process would manifest itself observationally as 
positive accelerations for blue-shifted high-velocity 
components and as negative accelerations for red-shifted high-velocity components. Different spiral theories are schematically represented in Figure~\ref{f:spiralcomp}.

\subsubsection{Spiral Density Wave Model \citep{Maoz1995}}

\citet{Maoz1995} argues that 
(i) maser components are periodic in disk radius with a characteristic wavelength 
of $\lambda_{char}\sim$0.75~mas; and  (ii) masers are located within 
several degrees of the disk midline. Point (ii) is required to avoid a wide range of 
accelerations of the masers due to spiral arms, which are not observed. If masers are 
near the midline, then the non-circular (peculiar) motion is largely 
perpendicular to the arms \citep{Toomre1981} and perpendicular 
to the line-of-sight (and therefore undetectable in Doppler shifts). \citet{Maoz1995} 
estimates that the component of the peculiar motion along the line-of-sight for high-velocity emission would 
not exceed 4.5 \velunitns, comparable to the measured 
accuracy of the Keplerian rotation curve \citep{Maoz1995}.

An apparent periodicity in the distribution of maser sky positions is obvious from VLBI 
images of the data. For formal quantification, we performed an 
analysis using the Lomb-Scargle periodogram for the 18 epochs of VLBI data 
described in Paper I.  First, we created a ``binary'' dataset from the VLBI 
positions, in which any disk position that had detected emission was assigned a ``flux density'' 
of ``1'', to ensure that disparity in flux density between red and blue-shifted emission 
did not affect the outcome of the analysis. We gridded the binary dataset 
at intervals of 0.005~mas, i.e., on scales much shorter than any expected $\lambda_{char}$, and 
placed zeroes at positions at which we did not detect, emission 
in the VLBI maps. Using the periodogram, we found a dominant period 
in the binary distribution of high-velocity 
maser positions of 0.75~mas (Figure~\ref{f:periodogram}), at the same value as 
that estimated by \citep{Maoz1995}.  We also predicted that emission 
should exist at a velocity of about $-$330 \velunits in the blue-shifted spectrum. 
In subsequent observations obtained with the GBT on 2003 October 23, the emission predicted 
by the periodogram analysis was detected at $-$329 \velunitns. 
At $< $10 mJy, this emission would not have been detected in our VLBI observations.
  
We note that the model of \citet{Maoz1995} has densities in the spiral arms that can produce
maser emission, but not in the inter-arm regions. In the denser spiral arms, gas condensations 
naturally form. 
In this event, discrete maser-emitting condensations could be
tracked as they move across an arm.
They are not necessarily persistent in the
inter-arm region, i.e. discrete maser condensations may only exist
within arms and disperse in the inter-arm regions. 
As a result, clumps responsible for low-velocity emission do not survive to beam
high-velocity emission toward the observer after one quarter of a rotation, and
conversely, high-velocity clumps do not survive to eventually be seen as low-velocity
clumps.

\subsubsection{Spiral Shock Model (MM98)}

The specific predictions of MM98 for masers, in addition to periodicity 
in maser sky positions, are that (i) blue-shifted 
components are accelerating and they are located behind the disk midline, and 
(ii) red-shifted components are decelerating and should be located in front 
of the midline. By contrast, in a disk dominated by Keplerian dynamics, masers 
in front of the midline should have a positive sign of acceleration, and masers 
located behind the midline should have a negative sign of acceleration.

We do not find that the accelerations of high-velocity components conform to 
the MM98 model.  Both the red- and blue-shifted component accelerations are 
statistically near zero (see Section~\ref{ss:sect4.1} and Tables~2 and~3), although there is a 2.6$\sigma$ 
bias of the latter to negative accelerations which is not in agreement with predictions of MM98. 
We note that \citet{Yamauchi2005} 
presented evidence in support of MM98. However, \citet{Yamauchi2005} measured the 
acceleration of one blue-shifted component only, which had a positive acceleration 
at the 2$\sigma$ significance level. We obtain a negative acceleration for the 
same component (see Section~\ref{s:sect5}) when it is decomposed into 3 features. 
Of the eight blue-shifted components for which 
we measured accelerations, only one is positive.                   
\citet{Bragg2000} also does not find accelerations that would support the MM98 theory.

\subsubsection{Other Origins for the Periodicity}

We also considered the possibility that orbiting objects (e.g., stars) 
create the sub-structure in the NGC~4258 accretion disk, with for example some resonance
mechanism causing regularity in the spacings.  If the stellar density is the 
same in the central regions of NGC~4258 as in the center of the Milky Way, then integrating 
$\rho_* (R) = 1.2 \times 10^6 (R/0.425)^{-\alpha}$
between $R=$0.14 and 0.28~pc and assuming $\alpha$=1.4 \citep[from][]{Genzel2003} yields a 
stellar mass of  2$\times$10$^{4}$ M$_{\odot}$ in the spherical volume containing the maser disk, 
so there are likely at least 
several thousand stars. 
The cumulative effect of many passages of stars around the black hole might bring them 
into a circular orbit co-rotating with the disk, although we note that highly eccentric 
stellar orbits are found in our Galactic Center. Each star could then either open up a 
gap in the disk (if the disk is thinner than the star's Hill radius and the viscosity 
in the disk is sufficiently low) or accrete material from the disk. The vertical thickness 
of the maser disk is measured to be 12~$\mu$as (Paper I). We therefore estimate the lower 
limit on the mass of a star (or stellar cluster) required to create a gap completely 
through the disk to be $>$ 0.05 M$_{\odot}$ ($2r_{Hill} \approx 2 r_{orb}(m_*/3M_{bh})^{1/3} = 0.0003$ pc for
$\overline{r_{orb}}$=0.2 pc), such that any star, even a brown dwarf, 
could create a gap in the disk equal to its thickness. We note that a number of 
massive He I stars (of mass 30 to 100 M$_\odot$) exist between 0.04 to 0.5~pc (1$''$ to 12$''$) of the 
Galactic center \citep{Genzel2003}, and that a corresponding \about100 M$_{\odot}$ star 
in the NGC~4258 disk would create a gap of $2r_{hill/roche}$=0.006 pc. The stars would be at radii several 
orders of magnitude greater distances from the SMBH than their tidal disruption radii $r_{tidal} \simeq r_* (M_{bh}/m_*)^{1/3}$, where $r_{*}$ and m$_{*}$ are the radius and mass of the star respectively \citep[see e.g,][]{Bogdanovic2004}.

\section{Implications for the Accretion Disk from Low-Velocity Emission}
\label{s:implications_lv}

The range of accelerations measured for low-velocity components is consistent
with the warped disk model of \citet{Herrnstein2005} (see our Figures~\ref{f:herrndisk} and~\ref{f:velcontours}). 
However, the systematic trend in component acceleration with Doppler velocity 
(see Section~\ref{ss:sect4.2}) is not
predicted by this model. The trend is likely connected to some aspect of the
warped disk geometry that results in a preferred locus in disk radius and azimuth
angle within which low-velocity maser action is favored (Figure~\ref{f:threepanel}a) as a result of
geometry and orientation to the line of sight, and we
investigate this point more in the next paper of this series, in which we 
report new modeling of the 3D disk structure and dynamics based on the expanded dataset
presented in Paper I and in this paper. In order to 
understand other types of ``higher-order'' phenomena that could give rise to the trend
however, 
we attempted to reproduce the observations by (i) assuming circular orbits in the disk
with a spiral arm in the low-velocity maser region (Figure~\ref{f:threepanel}b) and (ii) 
allowing for eccentricity in maser orbits (Figure~\ref{f:threepanel}c). 
In the preliminary 
investigations that follow, we have assumed that the maser disk is flat and is viewed edge-on.

\subsection{Spiral Structure}

In order to investigate whether spiral structure could give rise to
the gradient evident in accelerations of low-velocity components with Doppler velocity, 
we performed $N$-body simulations for a maser orbiting a 
supermassive black hole \citep[$3.8 \times 10^{7}$ M$_{\odot}$;][]{Herrnstein2005} 
in a flat disk, and encountering a spiral arm of various masses (Figure~\ref{f:spiralacc}). 
For the simulations, we used a direct $N$-body integration code, NBODY0 \citep{Aarseth1985}, 
in which each particle is followed with its own integration step to account 
for a range of dynamical rates. See \citet{Aarseth1985} for more details.

We represented a spiral arm at $t= $0 (Figure~\ref{f:spiralacc}a) using 8000 particles 
distributed along an arbitrarily-chosen trailing logarithmic spiral arm of 
$r=r_0\exp[\,b(\phi-\phi_0)\,]$,  in which $b=$0.42, $\phi_0=$0\degrs 
and $r_{0}=$ 1.5~mas or 0.06~pc, with a radial extent of 10$^{-2}$~pc. 
The pattern speed of the arm was set to be half that of the Keplerian orbital velocity 
(by reducing the force on arm particles by a factor of 4) at any 
given radius, and gravitational interaction between arm particles was ignored. We based
values for the fractional radial extent and pattern speed of the arm roughly on those of
Galactic spiral arms. To avoid potentially infinite acceleration between arm particles
and the maser, and to better represent a smooth distribution of mass in the arm, we included a ``softening'' term of $5 \times 10^{-3}$ pc that was added in quadrature
to each maser-particle separation.

We varied the mass contained in the arm (bounded by the upper
mass limit for the maser disk determined by \citet{Herrnstein2005} of 
M$_{disk,upper}=$9$\times$10$^5$ M$_{\odot}$) for different runs of the code. 
For any given run, we followed the evolution
of a maser cloud through the spiral arm (Figure~\ref{f:spiralacc}a-c) for several hundred years,
sampling every 0.55 years.

The computed line-of-sight accelerations for the maser
showed significant deviations from that of a mass-less arm, for arm masses greater than a
few percent level of M$_{disk,upper}$ (Figure~\ref{f:spiralacc}d). We found the ``amplitude'' 
of the acceleration deviation to be directly proportional to arm mass and could 
reproduce observed trends in acceleration (see Section~\ref{ss:sect4.2}) for an arm mass of 15\% of the upper limit 
maser disk mass. In the simulations, the acceleration trend is reproduced when
the maser is ``in'' the spiral arm (between 100 and 200 years in Figure~\ref{f:spiralacc})
We note that our calculations include gravitational effects only and
do not attempt to incorporate shocks or magneto-hydrodynamic effects of spiral density wave theory.

\subsection{Eccentric Accretion Flows}

While accretion flows are generally believed to circularize on timescales 
much shorter than the viscous timescale, \citet{Statler2001} showed that 
elliptical orbits of $e$ $\le$ 0.3 can be long-lived if the disk is thin,   
and if the orbits are nested and confocal, with precession 
rates that maintain the alignment. 

We describe a system of nested elliptical orbits of identical eccentricity, $e$, of 
differing semi-major axes, $a_{semi}$, and with the same periapsis angle, $\omega$, about
a supermassive black hole.
The periapsis angle is defined here as the angle between the sky plane to the periastron point on each orbit
(increasing from the ``red-shifted'' 
midline in the sense of disk rotation). In the discussion that follows, it is important 
to note that no matter the $e$ and $\omega$ of the system, red-shifted high-velocity 
maser emission components must originate from an azimuthal angle of $\phi$ $\approx$ 0\degrns 
 (where $\phi-\omega$ is the angle of the component from periastron), and blue-shifted components from 
$\phi$ $\approx$ 180\degrns on the basis of velocity coherence and acceleration arguments. 

We first used the symmetry of the position-velocity (P-V) diagram of high-velocity emission to 
place limits on eccentricity.  At any given radius in the disk in the plane of the sky (i.e.,
where the radial orbital velocity component is perpendicular to the LOS), the Doppler velocity is given by

\begin{equation}
v_{los}=\left(\frac{GM}{r}\right)^{1/2}\left[1+e\cos(\phi-\omega)\right]^{1/2}\cos\phi+v_{sys}
\end{equation}

\noindent where $r$ is angular radius in the disk and $M=M_{BH}/D$. 
We can therefore compute the ratio of $v_{red,\,los}/v_{blue,\,los}$ for 
closed orbit ($e$ $<$ 1) values of $e$ and $\omega$ using $\phi$=0\degrs for the red-shifted high-velocity emission 
and $\phi$=180\degrs for the blue-shifted high-velocity emission. 
Note that this ratio is always unity for circular orbits 
($e=$ 0)  and for orbits of any eccentricity viewed along the semi-major axes i.e., for $\omega$=90\degrs or 270\degrs in our definition.
Using values of v$_{sys}$=472$\pm$4 \velunits 
\citep{Cecil1992} and a disk dynamical center position of $x_{0}$ = $-$0.1$\pm$0.1 mas 
estimated from jet continuum
data \citep{Herrnstein1997} within the 1$\sigma$ uncertainties (noting that, by allowing a 1$\sigma$ deviation in
each quantity we are actually using a combined uncertainty $> $1$\sigma$), we computed the 
{\it observed} ratio of $v_{red,\,los}/v_{blue,\,los}$ from our VLBI data to lie between 0.98 to 1.03. 
This constrains maser orbits to have either a low eccentricity of $<$0.05 or to be more highly-eccentric 
but viewed (nearly) along  the semi-major axes of the orbits ($\omega$=90\degrs or
270\degrns). The parameter space we {\it eliminate} using the P-V diagram is marked by
the white hatching of positive gradient in Figure~\ref{f:stretch}.

To constrain the eccentricity and periapsis angle further, we can also use the low-velocity acceleration 
data as a function of Doppler velocity. Instantaneous acceleration is given by
$\dot{v}_{los}= - (GM/r^2)\sin\phi$ and $r= a_{semi}(1 - e^2)/(1 + e \cos (\phi - \omega))$ 
so we can write

\begin{equation}
\frac{d\dot{v}_{los}}{dv_{los}} = \left[\frac{GM}{(1-e^2)^4 a_{semi}^3}\right]^\frac{1}{2}
\frac{\gamma^2\cos\phi - 2 \gamma e \sin(\phi-\omega)\sin\gamma}{\sin\phi}
\end{equation}

\noindent where $\gamma = 1 + e \cos (\phi - \omega)$ and  where we assume that low-velocity masers occupy the 
same orbit such that $a_{semi}$ is
a constant, implied by the restriction of low-velocity masers to a narrow portion of the P-V diagram.
Using the sign of the gradient, we can further exclude all periapsis angles of  90\degrs $\le$ $\omega$ $\le$ 270\degrs, 
marked by the white hatching of negative gradient in Figure~\ref{f:stretch}. Since the $e$ - $\omega$ parameter space
remaining after these eliminations is very small (regions with no hatching at all in Figure~\ref{f:stretch}), 
it is unlikely that the maser orbits are highly eccentric and observed along a special viewpoint.
However, this is a topic we revisit in the next paper of this series, in which we perform 3D accretion disk modeling
of the masers including eccentric maser orbits.

\section{Conclusions}

In this paper, we measured centripetal accelerations of maser spectral components for data 
spanning 1994 to 2004.
We found that high-velocity emission accelerations lie in the range %14
$-$0.7 to $+$0.7 \accunitns, indicating the emission originates within 13\degrs of the disk midline for
material in Keplerian rotation (see Tables 2 and 3). 
The good agreement of maser projected vertical positions ($y$-positions) of high-velocity emission derived
from the acceleration data, with those of VLBI images, confirms that   
masers trace true gas dynamics of the disk.
While the high-velocity accelerations do not support the MM98  model of trailing 
shocks associated with spiral arms in the disk, we find a spatial periodicity in 
high-velocity emission of wavelength 0.75~mas. This supports the model of \citet{Maoz1995} of spiral structure due to 
density waves in the disk. 

We measured accelerations of low-velocity emission in the range  %15 
7.7 to 8.9 \accunitns, which is consistent with emission that originates 
from a concavity in the front-side of the disk reported by \citet{Herrnstein2005}. 
We confirm a systematic trend in accelerations of low-velocity emission as a function of component Doppler velocity, 
found by \citet{Haschick1994} and \citet{Greenhill1995b}. 
Preliminary investigations into the origin of the trend suggest that eccentricity in maser orbits 
is unlikely to be the cause. The trend may be caused either by the effect of a stationary or
slowly moving spiral arm, or by a disk feature which causes a stationary pattern of radius versus
azimuth angle.

We are grateful to Maryam Modjaz and Paul Kondratko for
providing their GBT data. We thank Alar Toomre for helpful discussions.

\bibliographystyle{apj}
\bibliography{newrefs2}

\clearpage


%% the file, uncomment the \plotone, \plottwo, and
%% If you need a layout that cannot be achieved with \plotone or
%% \plottwo, you can invoke the graphicx package directly with the

%% \includegraphics command or use \plotfiddle. For more information,
%% please see the tutorial on "Using Electronic Art with AASTeX" in the
%% documentation section at the AASTeX Web site,
%% http://www.journals.uchicago.edu/AAS/AASTeX.
%%
%% The examples below also include sample markup for submission of
%% supplemental electronic materials. As always, be sure to check

%% the instructions to authors for the journal you are submitting to
%% for specific submissions guidelines as they vary from

%% This example uses \plotone to include an EPS file scaled to
%% 80% of its natural size with \epsscale. Its caption
%% has been written to indicate that additional figure parts will be
%% available in the electronic journal.

\clearpage

\begin{deluxetable}{rlrllccl}
\tabletypesize{\scriptsize}
%\rotate
\tablewidth{0pt}
\tablecaption{The Dataset.\label{t:dataset}}
\tablehead{
\multicolumn{3}{c}{{\bf--------  Epoch --------}} & 
\multicolumn{2}{c}{{\bf Observations }}           &
\multicolumn{2}{c}{{\bf Observing Details }}      &
\multicolumn{1}{l}{{\bf Comments}}               \\
\multicolumn{1}{c}{No.}                           &
\multicolumn{1}{c}{Date}                          & 
\multicolumn{1}{c}{Day}                           &
\multicolumn{1}{c}{Program}                          &
\multicolumn{1}{c}{Telescope}                     & 
\multicolumn{1}{c}{$\Delta$v\tablenotemark{a}}   &
\multicolumn{1}{c}{Sensitivity\tablenotemark{a}}  & 
\colhead{}                                       \\
\colhead{}                                        & 
\colhead{}                                        &
\multicolumn{1}{c}{No.}                           &
\multicolumn{1}{c}{Code}                          &
\colhead{}                                        & 
\multicolumn{1}{c}{(\velunitns)}                    &
\multicolumn{1}{c}{(mJy)}                         & 
\colhead{}} 
\startdata
 1  & 1994 Apr 19  &    0 & BM19    & VLBA   & 0.21  & 43    & Systemic, red and  blue-shifted emission \\
 2  & 1995 Jan 07  &  263 & AG448 & VLA-CD & 0.33  & 25    & Systemic, red and  blue-shifted emission \\
 3  & 1995 Jan 08  &  264 & BM36a   & VLBA   & 0.21  & 80    & Systemic, red and  blue-shifted emission\\
 4  & 1995 Feb 23  &  310 & AG448 & VLA-D  & 0.33  & 25    & Systemic, red and  blue-shifted emission \\
 5  & 1995 Mar 16  &  331 & \nodata & EFLS   & 0.33  & 85    & Red-Shifted emission only\\
 6  & 1995 Mar 24  &  339 & \nodata & EFLS   & 0.33  & 90    & Red-Shifted emission only\\
 7  & 1995 Mar 25  &  340 & \nodata & EFLS   & 0.33  & 95    & Red-Shifted emission only\\
 8  & 1995 Apr 04  &  350 & \nodata & EFLS   & 0.33  & 85    & Red-Shifted emission only\\
 9  & 1995 Apr 20  &  366 & AG448 & VLA-D  & 0.33  & 20    & Systemic, red and  blue-shifted emission \\
 10 & 1995 May 29  &  405 & BM36b   & VLBA   & 0.21  & 30    & Systemic, red and  blue-shifted emission \\
 11 & 1995 Jun 08  &  415 & AG448 & VLA-AD & 0.33  & 20 & Systemic, red and  blue-shifted emission    \\
 12 & 1995 Jun 25  &  432 & \nodata & EFLS   & 0.33  & 110   & Red-Shifted emission only  \\
 13 & 1995 Jul 29  &  466 & AG448 & VLA-A  & 0.33  &\nodata& No data\\
 14 & 1995 Sep 09  &  508 & AG448 & VLA-AB & 0.33  &\nodata& No data \\
 15 & 1995 Nov 09  &  569 & AG448 & VLA-B  & 0.33  & 37-60 & Systemic, red and  blue-shifted emission   \\
 16 & 1996 Jan 11  &  632 & AG448 & VLA-BC & 0.33  & 20    & Systemic, red and  blue-shifted emission \\
 17 & 1996 Feb 22  &  674 & BM56a   & VLBA   & 0.21  & 25    & Systemic, red and  blue-shifted emission\\    	
 18 & 1996 Feb 26  &  678 & AG448 & VLA-C  & 0.33  & 23-30 & Systemic, red and  blue-shifted emission  \\
 19 & 1996 Mar 29  &  710 & AG448 & VLA-C  & 0.33  & 20    & Systemic, red and  blue-shifted emission \\           
 20 & 1996 May 10  &  752 & AG448 & VLA-CD & 0.33  & 15-24 & Systemic, red and  blue-shifted emission\\
 21 & 1996 Jun 27  &  799 & AG448 & VLA-D  & 0.33  &\nodata& No data\\
 22 & 1996 Aug 12  &  846 & AG448 & VLA-D  & 0.33  & 23-36 & Systemic, red and  blue-shifted emission\\
 23 & 1996 Sep 21  &  886 & BM56b   & VLBA   & 0.21  & 20    & High-velocity emission only\\
 24 & 1996 Oct 03  &  898 & AG448 & VLA-AD & 0.33  & 50    & Systemic, red and  blue-shifted emission   \\
 25 & 1996 Nov 21  &  947 & AG448 & VLA-A  & 0.33  & 21-26 & Systemic and red-shifted emission \\
 26 & 1997 Jan 20  & 1007 & AG448 & VLA-AB & 0.33  & 27-37 & Systemic and red-shifted emission  \\
 27 & 1997 Feb 10  & 1028 & AG448 & VLA-AB & 0.33  & 17-23 & Systemic and red-shifted emission\\  
 28 & 1997 Mar 06  & 1052 & BM56c   & VLBA   & 0.21  & 4.7   & Systemic, red and  blue-shifted emission  \\
 29 & 1997 Oct 01  & 1261 & BM81a   & VLBA   & 0.21  & 4.1   & Systemic, red and  blue-shifted emission  \\
 30 & 1998 Jan 27  & 1379 & BM81b   & VLBA   & 0.21  & 5.0   & Systemic, red and  blue-shifted emission   \\
 31 & 1998 Sep 05  & 1600 & BM112a  & VLBA   & 0.21  & 4.6   & Systemic, red and  blue-shifted emission\\
 32 & 1998 Oct 18  & 1643 & BM112b  & VLBA   & 0.42  & 4.6   & Systemic and red-shifted emission         \\    
 33 & 1998 Nov 16  & 1672 & BM112c  & VLBA   & 0.42  & 4.4   & Systemic and blue-shifted emission\\ 
 34 & 1998 Dec 24  & 1710     & BM112d  & VLBA   & 0.42  &\nodata& No data \\
 35 & 1999 Jan 28  & 1745 & BM112e  & VLBA   & 0.42  & 4.8   & Systemic and blue-shifted emission  \\     
 36 & 1999 Mar 19  & 1795 & BM112f  & VLBA   & 0.42  & 4.5   & Systemic and red-shifted emission    \\      
 37 & 1999 May 18  & 1855 & BM112g  & VLBA   & 0.42  & 5.2   & Systemic and blue-shifted emission   \\
 38 & 1999 May 26  & 1863 & BM112h  & VLBA   & 0.21  & 3.5   & Systemic, red and  blue-shifted emission    \\
 39 & 1999 Jul 15  & 1913     & BM112i  & VLBA   & 0.42  &\nodata& No data \\
 40 & 1999 Sep 15  & 1975 & BM112j  & VLBA   & 0.42  & 6.3   & Systemic and blue-shifted emission       \\
 41 & 1999 Oct 29  & 2019 & BM112k  & VLBA   & 0.42  & 4.7   & Systemic and blue-shifted emission          \\
 42 & 2000 Jan 07  & 2089 & BM112l  & VLBA   & 0.42  & 3.9   & Systemic and blue-shifted emission         \\
 43 & 2000 Jan 30  & 2112 & BM112m  & VLBA   & 0.42  & 4.7   & Systemic and red-shifted emission    \\ 
 44 & 2000 Mar 04  & 2146 & BM112n  & VLBA   & 0.42  & 4.3   & Systemic and blue-shifted emission          \\
 45 & 2000 Apr 12  & 2185 & BM112o  & VLBA   & 0.42  & 5.5   & Systemic and red-shifted emission  \\
 46 & 2000 May 04  & 2207 & BM112p  & VLBA   & 0.42  & 5.0   & Systemic and blue-shifted emission   \\
 47 & 2000 Aug 12  & 2307 & BG107   & VLBA   & 0.42  & 7.2   & Systemic, red and  blue-shifted emission \\
 48 & 2003 Apr 10  & 3298 & \nodata & GBT    & 0.21     & 2.9      & Systemic, red and  blue-shifted emission \\ 
 49 & 2003 Oct 23  & 3474 & \nodata & GBT    & 0.21  & 2.9   & Systemic, red and  blue-shifted emission \\ 
 50 & 2003 Dec 08  & 3520 & \nodata & GBT    & 0.21  & 2.9   & Systemic, red and  blue-shifted emission \\ 
 51 & 2004 May 21  & 3685 & AH847 & VLA    & 0.21    &  20     &   Portions of red-shifted spectrum                                      \\ 
\enddata
\tablenotetext{a}{Channel spacing  and sensitivities (r.m.s)
for epochs 1 -- 27 taken from \citet{Bragg2000}; 48 -- 47 from Paper I and
for GBT data (M. Modjaz, P. Kondratko, private communications)}
%% You can append references to a table using the \tablerefs command.
\end{deluxetable}

\clearpage

\begin{deluxetable}{rrrcccc} 
\tabletypesize{\scriptsize}
%\rotate
\tablewidth{0pt}
\tablecaption{Fitted Accelerations For Red-Shifted High-Velocity Doppler Components.
\label{t:redtable}}
\tablehead{
\multicolumn{3}{c}{{\bf -------------- Component --------------}}&
\multicolumn{2}{c}{{\bf --------- Epochs ---------} }&
\multicolumn{1}{c}{{\bf Time-Averaged}}&
\multicolumn{1}{c}{{\bf Component}}\\
%\multicolumn{2}{c}{}\\
\multicolumn{1}{c}{No.} &
\multicolumn{1}{c}{{\bf Velocity\tablenotemark{a}}}&
\multicolumn{1}{c}{{\bf Acceleration\tablenotemark{b}}}& 
\multicolumn{1}{c}{No. in Fit} &
\multicolumn{1}{c}{Time Baseline}&
\multicolumn{1}{c}{{\bf Azimuth Angle}}&
\multicolumn{1}{c}{{\bf Radius}\tablenotemark{c}}\\
%\colhead{} &
\colhead{} &
%\colhead{}                    &
\multicolumn{1}{c}{{\bf (\velunitns)}}&   
\multicolumn{1}{c}{{\bf (\accunitns)}}& 
\colhead{} &
\multicolumn{1}{c}{(yr)}&
\multicolumn{1}{c}{(deg)}&    
\multicolumn{1}{c}{(pc)}} 
\startdata
1&1248.14$\pm$0.02  &  -0.06 $\pm$0.02 & 24  &   6.0                    & -1.6$\pm$0.6 &  0.28\\
2&1250.80$\pm$0.03  &  -0.35 $\pm$0.03 & 24  &   6.0                    & -8.9$\pm$0.8 &  0.28\\
3&1252.24$\pm$0.01  &   0.17 $\pm$0.01 & 24  &   6.0                    &  4.3$\pm$0.3 &  0.28\\
4&1254.34$\pm$0.02  &  -0.01 $\pm$0.03 & 24  &   6.0                    & -0.3$\pm$0.7 &  0.28\\
5&1270.89$\pm$0.05  &   0.11 $\pm$0.04 & 24  &   6.0                    &  2.5$\pm$1.0 &  0.27\\
6&1282.21$\pm$0.06  &  -0.15 $\pm$0.08 &  4  &   0.7                    & -3.1$\pm$1.8 &  0.26\\
7&1283.87$\pm$0.08  &  -0.05 $\pm$0.14 &  4  &   0.7                    & -1.1$\pm$3.0 &  0.26\\
8&1309.08$\pm$0.01  &  -0.09 $\pm$0.01 & 30  &   9.0                    & -1.7$\pm$0.1 &  0.24\\
9&1328.64$\pm$0.04  &   0.73 $\pm$0.02 & 29  &   5.3                    & 12.6$\pm$0.6 &  0.23\\
10&1330.73$\pm$0.06  &   0.73 $\pm$0.03 & 29  &   5.3                    & 12.4$\pm$0.7 &  0.23\\
11&1337.36$\pm$0.04  &  -0.16 $\pm$0.03 & 35  &   6.3                    & -2.7$\pm$0.5 &  0.23\\
12&1339.58$\pm$0.01  &  -0.28 $\pm$0.01 & 35  &   6.3                    & -4.6$\pm$0.3 &  0.22\\
13&1351.00$\pm$0.03  &   0.40 $\pm$0.03 & 34  &   6.0                    &  6.1$\pm$0.5 &  0.22\\
14&1353.85$\pm$0.03  &   0.11 $\pm$0.03 & 34  &   6.0                    &  1.8$\pm$0.4 &  0.22\\
15&1355.56$\pm$0.02  &  -0.10 $\pm$0.02 & 34  &   6.0                    & -1.5$\pm$0.3 &  0.22\\
16&1395.59$\pm$0.09  &   0.24 $\pm$0.05 & 21  &   4.3                    &  3.1$\pm$0.7 &  0.20\\
17&1398.09$\pm$0.05  &   0.58 $\pm$0.03 & 21  &   4.3                    &  7.3$\pm$0.4 &  0.20\\
18&1403.60$\pm$0.01  &  -0.05 $\pm$0.02 & 34  &   6.0                    & -0.6$\pm$0.2 &  0.19\\
19&1406.74$\pm$0.05  &  -0.06 $\pm$0.04 & 34  &   6.0                    & -0.8$\pm$0.5 &  0.19\\
20&1449.43$\pm$0.04  &   0.30 $\pm$0.04 & 25  &   2.9                    &  3.0$\pm$0.5 &  0.18\\
21&1452.35$\pm$0.01  &  -0.40 $\pm$0.02 & 25  &   2.9                    & -4.0$\pm$0.2 &  0.18\\
22&1452.80$\pm$0.14  &   0.04 $\pm$0.06 & 25  &   2.9                    &  0.4$\pm$0.6 &  0.18\\
23&1455.26$\pm$0.19  &  -0.33 $\pm$0.09 & 25  &   2.9                    & -3.2$\pm$0.9 &  0.17\\
24&1468.08$\pm$0.07  &   0.34 $\pm$0.08 &  4  &   2.6                    &  3.2$\pm$0.8 &  0.17\\
\hline
\multicolumn{2}{r}{{{\bf Mean Acceleration:}}\tablenotemark{d}}&\multicolumn{1}{r}{0.02 $\pm$0.06}&
\multicolumn{2}{c}{{{\bf Mean Azimuth Angle:}}\tablenotemark{d}}&\multicolumn{1}{c}{0.2$\pm$3.6}\\
\enddata
\tablenotetext{a}{Component velocities (relativistic definition) for fits that met requirements
of Section~X are quoted for 
1999 October 10, day 2000 of our monitoring campaign. Uncertainties are the 1-$\sigma$ errors scaled 
by $\chi^2$ per degree of freedom.} 
\tablenotetext{b}{Uncertainties are the 1-$\sigma$ errors scaled 
by $\chi^2$ per degree of freedom.}
\tablenotetext{c}{Calculated using a black hole mass of
3.8 $\times$ 10$^{7}$ M$_{\odot}$ \citep{Herrnstein2005} for a flat disk.}
\tablenotetext{d}{Quantities are the weighted mean and the weighted deviation 
from the mean.}
\label{t:table2}
\end{deluxetable}

%liznote: 1999 October 10 is JD 2451462.5 (day 2000 of monitoring)
% day 0 of monitoring is 2449462.5
% bm81b is on 1998 Jan 27 or JD 2450841.5
\clearpage

\begin{deluxetable}{rrrcccc} 
%\label{t:table3}
\tabletypesize{\scriptsize}
%\rotate
\tablewidth{0pt}
\tablecaption{Fitted Accelerations For Blue-Shifted High-Velocity Doppler Components.
\label{t:bluetable}}
\tablehead{
\multicolumn{3}{c}{{\bf -------------- Component --------------}}&
\multicolumn{2}{c}{{\bf --------- Epochs ---------} }&
\multicolumn{1}{c}{{\bf Time-Averaged}}&
\multicolumn{1}{c}{{\bf Component}}\\
%\multicolumn{2}{c}{}\\
\multicolumn{1}{c}{No.} &
\multicolumn{1}{c}{{\bf Velocity\tablenotemark{a,b}}}&
\multicolumn{1}{c}{{\bf Acceleration\tablenotemark{b}}}& 
\multicolumn{1}{c}{No. in Fit} &
\multicolumn{1}{c}{Time Baseline}&
\multicolumn{1}{c}{{\bf Azimuth Angle}}&
\multicolumn{1}{c}{{\bf Radius}\tablenotemark{c}}\\
%\colhead{} &
\colhead{} &
%\colhead{}                    &
\multicolumn{1}{c}{{\bf (\velunitns)}}&   
\multicolumn{1}{c}{{\bf (\accunitns)}}& 
\colhead{} &
\multicolumn{1}{c}{(yr)}&
\multicolumn{1}{c}{(deg)}&    
\multicolumn{1}{c}{(pc)}} 
\startdata
1& -282.21$\pm$0.07& -0.47 $\pm$0.10 & 6 & 1.5 & -13.2$\pm$2.9 & 0.29 \\
2& -284.00$\pm$0.06& -0.12 $\pm$0.08 & 6 & 1.5 & -3.3$\pm$2.3&  0.29 \\
3& -286.05$\pm$0.13& -0.38 $\pm$0.13 & 6 & 1.5 & -10.3$\pm$3.5&  0.29\\
4& -374.20$\pm$0.05& -0.32 $\pm$0.26 & 5 & 1.1 & -5.6$\pm$4.7& 0.23\\
5& -375.78$\pm$0.22& -0.72 $\pm$0.49 & 5 & 1.1 & -12.6$\pm$ 8.6& 0.23\\
6& -435.00$\pm$0.03& -0.29 $\pm$0.05 & 16& 3.5 & -3.9$\pm$ 0.7&  0.20\\
7& -439.97$\pm$0.02& +0.04 $\pm$0.04 & 15& 2.9 &0.5$\pm$ 0.5&  0.20\\
8& -514.48$\pm$0.04& -0.24 $\pm$0.16 & 4 & 0.5 &-2.3$\pm$ 1.5&  0.17\\
\hline
\multicolumn{2}{r}{{{\bf Mean Acceleration:}}\tablenotemark{d}}&\multicolumn{1}{r}{-0.21 $\pm$0.08}&
\multicolumn{2}{c}{{{\bf Mean Azimuth Angle:}}\tablenotemark{d}}&\multicolumn{1}{c}{-3.3$\pm$6.9}\\
\enddata
\tablenotetext{a}{Component velocities (relativistic definition) are quoted for 
1999 October 10, day 2,000 of our monitoring campaign. Uncertainties are the 1-$\sigma$ errors scaled 
by $\chi^2$ per degree of freedom.} 
\tablenotetext{b}{Uncertainties are the 1-$\sigma$ errors scaled 
by $\chi^2$ per degree of freedom.}
\tablenotetext{c}{Calculated using a black hole mass of
3.8 $\times$ 10$^{7}$ M$_{\odot}$ \citep{Herrnstein2005} for a flat disk.}
\tablenotetext{d}{Quantities are the weighted mean and the weighted deviation 
from the mean.}
\end{deluxetable}

\clearpage

\begin{deluxetable}{ccc} 
%\label{t:table4}
\tabletypesize{\scriptsize}
%\rotate
\tablewidth{0pt}
\tablecaption{Accelerations for Low-Velocity Maser Doppler Components.
\label{t:systable}}
\tablehead{
\multicolumn{1}{c}{{\bf Velocity\tablenotemark{a}}}&
\multicolumn{1}{c}{{\bf Acceleration\tablenotemark{b}}}&
\multicolumn{1}{c}{{\bf Radius in Geometric Model\tablenotemark{c}}}\\
\multicolumn{1}{c}{{\bf (\velunitns)}}&   
\multicolumn{1}{c}{{\bf (\accunitns)}}&
\multicolumn{1}{c}{{\bf (mas)}}}
%colhead{}
%multicolumn{1}{c}{(Yrs)}                  }
\startdata
  438&  7.65$\pm$0.15  &4.20$\pm$0.04\\
  448&  7.81$\pm$0.13  &4.16$\pm$0.03\\
  458&  7.85$\pm$0.13  &4.15$\pm$0.03\\
  468&  7.77$\pm$0.08 & 4.17$\pm$0.02\\
  478&  7.84$\pm$0.17  &4.15$\pm$0.04\\
  488&  8.29$\pm$0.12  &4.04$\pm$0.03\\
  498&  8.16$\pm$0.13  &4.07$\pm$0.03\\
  508&  8.09$\pm$0.09 & 4.09$\pm$0.02\\
  518&  8.75$\pm$0.09 & 3.93$\pm$0.02\\
  528&  8.55$\pm$0.06 & 3.97$\pm$0.02\\
  538&  8.36$\pm$0.10 & 4.02$\pm$0.02\\
  548&  8.87$\pm$0.27 & 3.89$\pm$0.05\\
\enddata
%% Any table notes must follow the \end{tabular} command.
\tablenotetext{a}{Data have been binned into 10 \velunits intervals.}
\tablenotetext{b}{Uncertainties are the standard deviation of the mean for each bin.}
\tablenotetext{c}{For the best-fitting disk geometry of \citet{Herrnstein2005}. See also
Figure~\ref{f:velcontours}.}
\end{deluxetable}

\clearpage

\begin{deluxetable}{ccccccc}
\label{t:acceleration}
\tabletypesize{\scriptsize}
%\tabletypesize{\small}
\rotate
\tablewidth{0pt}
\tablecaption{Studies of Acceleration Among NGC\,4258 Maser Components}
\tablehead{
\colhead{Study\tablenotemark{a}}         &
\colhead{Antennas\tablenotemark{b}}      & 
\colhead{Sensitivity\tablenotemark{c}}   &
\colhead{Dates}                          &
%\colhead{Frequency}                          &
\colhead{Velocity}& 
\colhead{Components}               &
\colhead{Acceleration}         \\
&
&
%&
\colhead{(mJy)}&
&
\colhead{Range\tablenotemark{d}}&
\colhead{Tracked}&
\colhead{Range}\\
&
&
&
&
\colhead{(km s$^{-1}$)}&
&
\colhead{(km s$^{-1}$ yr$^{-1}$)}\\
}

\startdata
\citet{Haschick1990} & HAYS & 55$-$70         & 7/1986$-$11/1989 & \ 340 to 590 & 1 (systemic)& 10  \\
                     &      &                 &$\approx$ monthly &              &   &    \\
                     &      &                 &                  &              &   &     \\          
\citet{Haschick1994} & HAYS &     30$-$55     & 7/1986$-$11/1993 & \ 340 to 590 & 4 (systemic) & 6.2$-$10.4   \\
                     &      &                 & $\approx$ monthly&              &   &    \\
                     &      &                 &                  &              &   &     \\
\citet{Greenhill1995b}&EFLS &  30             &10/1984$-$12/1986 & 335 to 665   & 12& 8.1$-$10.9 \\
                      &     &                 & every few days $-$ months & 1185 to 1515 & 20  &$<$$\pm$1\\
                      &     &                 &02/1993$-$03/1994& $-$570 to $-$240  & 3 &  $<$$\pm$1  \\
                      &     &                 & more frequently   &              &   &     \\
                      &     &                 &                   &              &   &     \\
\citet{Nakai1995}     & NRO &  13$-$20        &    1992           & $-$1197 to 2664 & 13 (systemic) & 7.2$-$11.1\\
                      &     &                 &  $\approx$ weekly &   \nodata              & 8 (red) & $<$$\pm$0.7\\
                      &     &                 &                   &   \nodata              & 1 (blue) &\\
                      &     &                 &                   &                 &       &\\
\citet{Herrnstein1999}&VLBA+VLA (3 epochs)&  6$-$12 & 04/1994$-$02/1996&430 to 590        & 30 (systemic)  & 6.8$-$11.6\\
                   & VLBA+VLA+GB (1 epoch)&         &                  &1225 to 1490      & 1 (red)   &   0.0  \\
                   &                      &         &                  &$-$450 to $-$350  & 0 (blue   & \\
                   &                      &         &                  &                  &      & \\
\citet{Bragg2000}& VLBA (5 epochs)        &  40     &04/1994$-$02/1997 &390 to 660        & 12 (systemic)  & 7.5$-$10.4   \\
                 & EFLS (5 epochs)        &  95     &                  &1235 to 1460      & 17  (red) & -0.77$-$0.38\\
                 & VLA (17 epochs)        &  20     &                  &$-$460 to $-$420,$-$390 to $-$350  & 2  (blue)  & -0.41,$+$0.04\\
                 &                        &         &                  &                                   &      & \\
\citet{Yamauchi2005}& NRO                 & 8$-$140 &01/1992$-$04/2005                 & $-$725 to 1720  & 14 (red)  &-0.52$-$0.41    \\
                    &                     &         & 30 irregularly spaced spectra    &    \nodata   & 1 (blue)  & +0.2 \\
                    &                     &         &                                  &    \nodata   &         &   \\

This Work           &   Bragg Dataset(27 epochs)   & 20$-$95    & 04/1994$-$05/2004          & -850 to 1850 & 25 (systemic)& 6.1$-$10.8 \\
                    &   VLBA+VLA+EFLS (18 epochs)  & 3.2$-$7.8  & on average every 2.5 months &\nodata & 24 (red) & -0.4$-$0.7 \\              
                    &   GBT(3 epochs)              & a few mJy  &                            &  \nodata& 8 (blue) & -0.7$-$0.04\\
                    &   VLA(1 epoch)               &     5      &                            &              &     & \\
\enddata

%% Any table notes must follow the \end{tabular} command.

\tablenotetext{a}{\citet{Haschick1990} report primarily on a variability study, but 
note that the 465 km s$^{-1}$ component is accelerating.  \citet{Haschick1994}, 
\citet{Greenhill1995b}, \citet{Nakai1995}, \citet{Bragg2000}, and \citet{Yamauchi2005} determine accelerations by 
tracking local spectral maxima and fitting velocity drifts as a linear 
function of time.  
\citet{Bragg2000} performs multiple-Gaussian component decomposition to high-velocity spectra, but
not to systemic spectra.
\citet{Herrnstein1999} determines accelerations by applying a 
Bayesian statistical analysis technique to all possible pairings of components 
among epochs.  Two of the three high velocity studies, i.e., \citet{Greenhill1995b} and 
\citet{Nakai1995}, obtain upper limits only. In this work, we perform multiple-Gaussian
component decomposition to high-velocity {\it and} systemic spectra in multi-epoch
simultaneous fits. 
}

\tablenotetext{b}{HAYS: Haystack 37-m antenna; EFLS: Effelsberg 100-m antenna; 
NRO: Nobeyama Radio Observatory 45-m; VLBA: Very Long Baseline Array; VLA: 
27$\times$25-m Very Large Array; GB: NRAO 140-ft.
}

\tablenotetext{c}{Noise level (1$\sigma$) per channel in emission-free portions of the 
spectra (single dish or VLA) or synthesis images (VLBA). See relevant publications
for further details.
}

\tablenotetext{d}{Ranges for \citet{Greenhill1995b} and \citet{Herrnstein1999} are 
approximate.  Ranges for \citet{Bragg2000} are for all epochs except the 
first.  The velocity ranges for the first epoch were all shifted by 20 km 
s$^{-1}$ towards higher velocities.  
}
\end{deluxetable}

\clearpage

\clearpage

\clearpage

\clearpage

\clearpage

\begin{figure}
\includegraphics[scale=0.8]{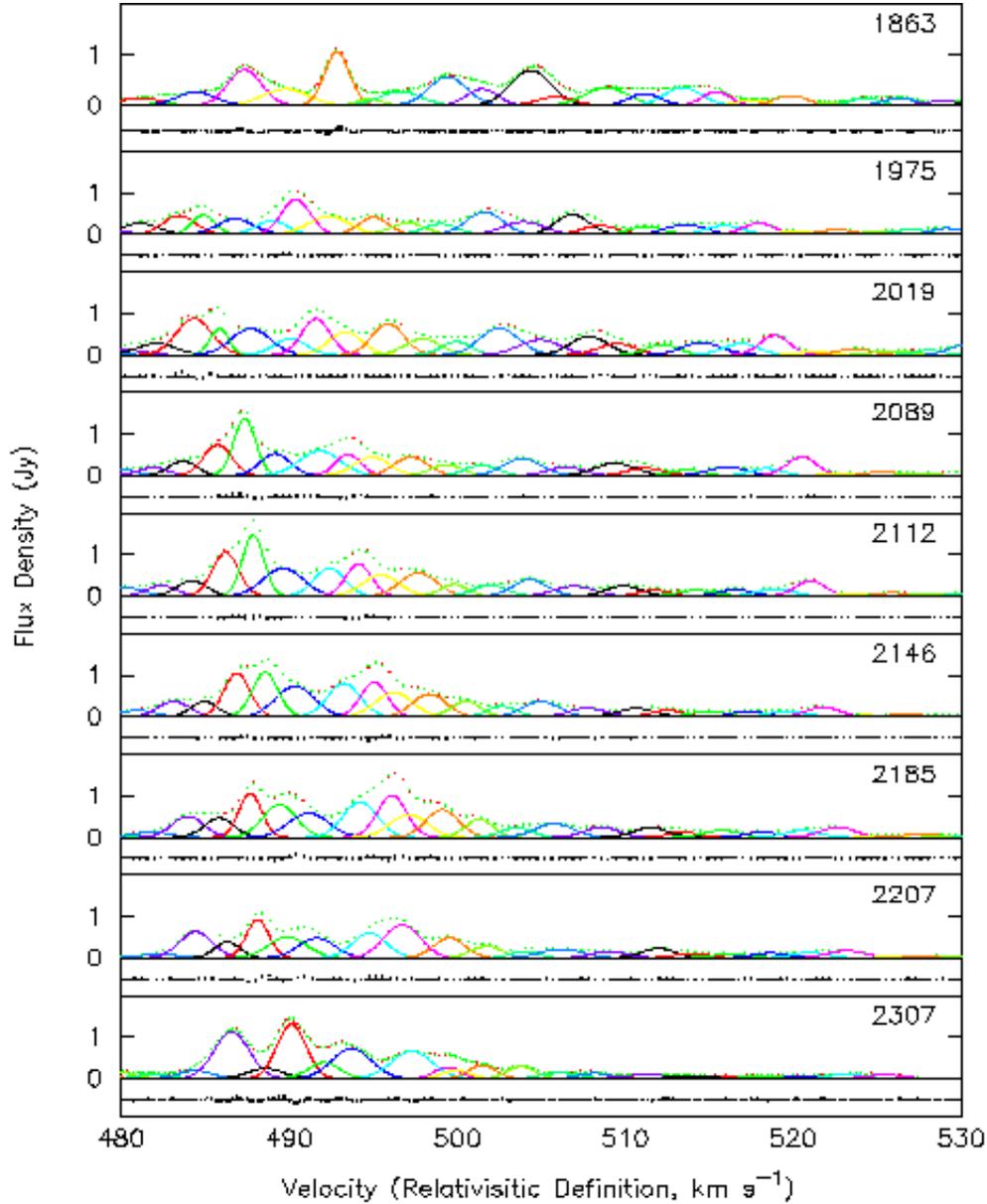}
\caption{Example of Gaussian decomposition for low-velocity emission. 
A velocity-portion of a fit over 9 epochs is
shown, in which dots are data and black symbols are
the residuals including 1$\sigma$ error bars. Individual model Gaussians are coded using 
the same color at each epoch, such that Doppler velocity drifts are clearly evident.
Numbers in the upper right corner of each panel are the number of days since 1994 April 19. 
\label{f:sysfit}}
\end{figure}

\clearpage

\begin{figure}
\includegraphics[angle=-90,scale=0.7]{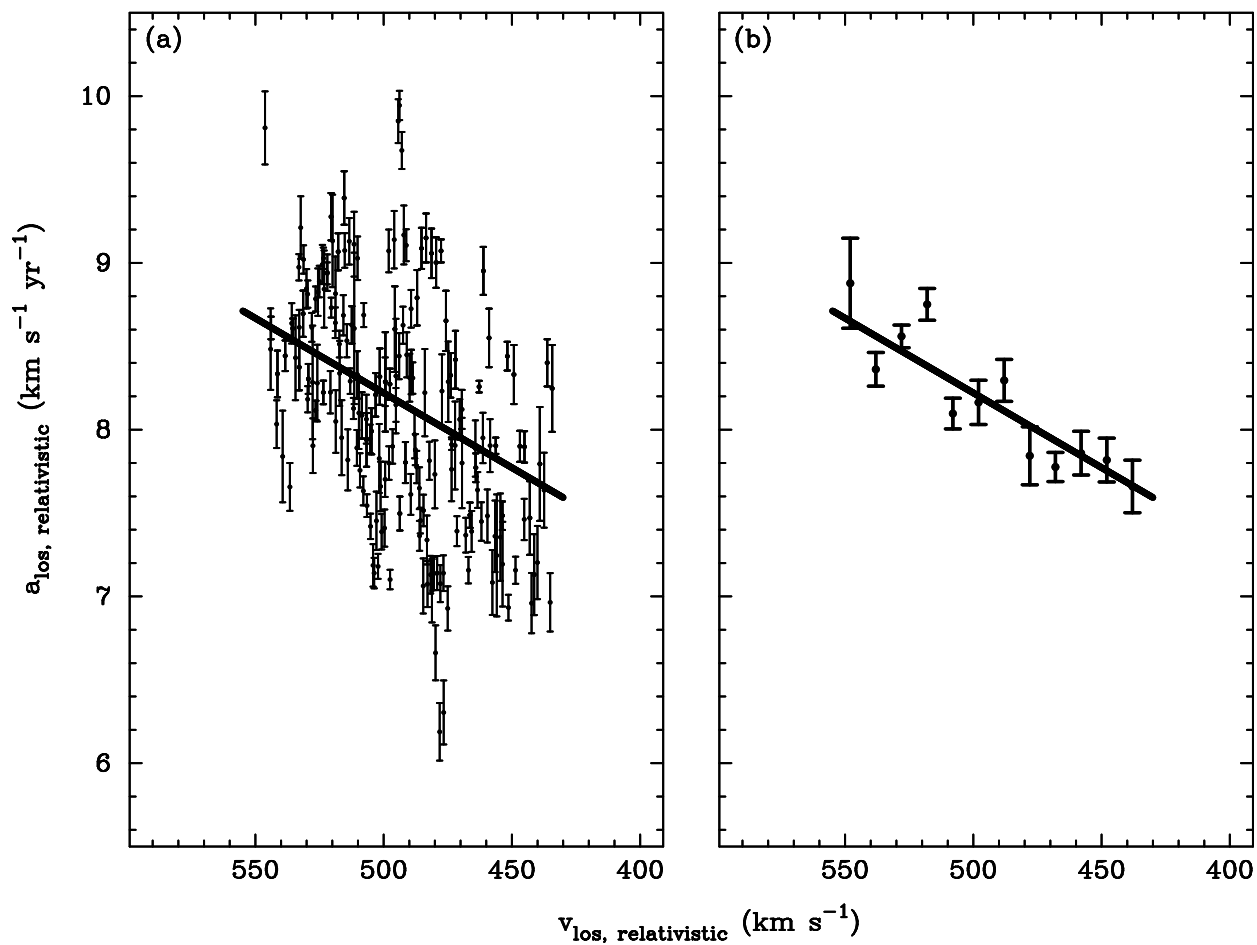}
%\includegraphics[scale=0.8]{showcrossing4paper.ps}
%\plotone{systemic_trend.ps}
\caption{Results of Gaussian decomposition for low-velocity emission.
{\it (a)} Measured accelerations from each of the four time-consecutive 9-epoch fits.
Each fit required $\sim$55 Gaussian components. The line-of-best-fit to all the data is shown. 
{\it (b)} Data in {\it (a)} binned into 10 \velunits intervals. The errorbars are the
standard deviation of the mean in each bin.
\label{f:systrend}} 
\end{figure}

\clearpage

\begin{figure}
\includegraphics[angle=-90,scale=0.7]{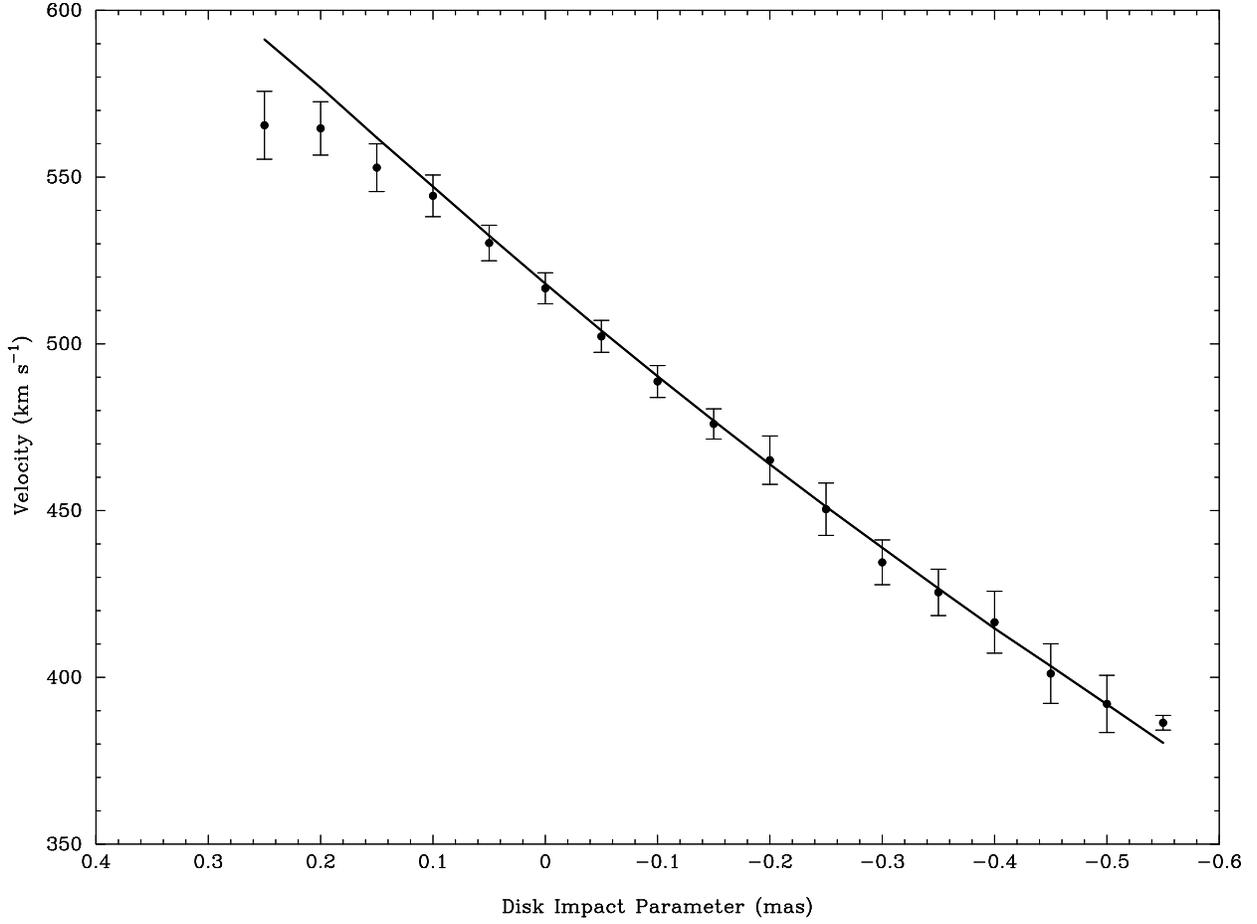}
%\includegraphics[scale=0.8]{showcrossing4paper.ps}
%\plotone{systemic_trend.ps}
\caption{Predicted position-velocity diagram for a
linear fit to the low-velocity acceleration data (see Figure~6). 
Symbols mark observed
data from the 18 VLBI epochs of Paper I, binned into 0.05 mas intervals.
The curve shows the diagram predicted from the line-of-best-fit to the
acceleration data. Curve and data are in close correspondence except
at three most positive values of impact parameter.
\label{f:systrendpv}} 
\end{figure}

\clearpage

\begin{figure}
\includegraphics[angle=-90,scale=0.7]{f8.ps}
\caption{Comparison of acceleration measurements for red-shifted maser spectral 
components: the current work (black circles); \citet{Bragg2000} (red triangles); 
\citet{Yamauchi2005} 1997 to 2005 (green squares); \citet{Yamauchi2005} 1992 to 2005 
(blue stars). \citet{Yamauchi2005} derive accelerations from their dataset over
two time ranges, one of which is a subset of the other.
All velocities have been converted to a relativistic definition, are referenced to the LSR, and 
have been adjusted to a common reference date (our monitoring day 2000). Measurements from the different 
studies are in broad agreement, however this work tends to identify more components in any given blend 
compared with earlier work.
\label{f:acccomp}}
\end{figure}

\clearpage

\begin{figure}
\includegraphics[angle=0,scale=0.7]{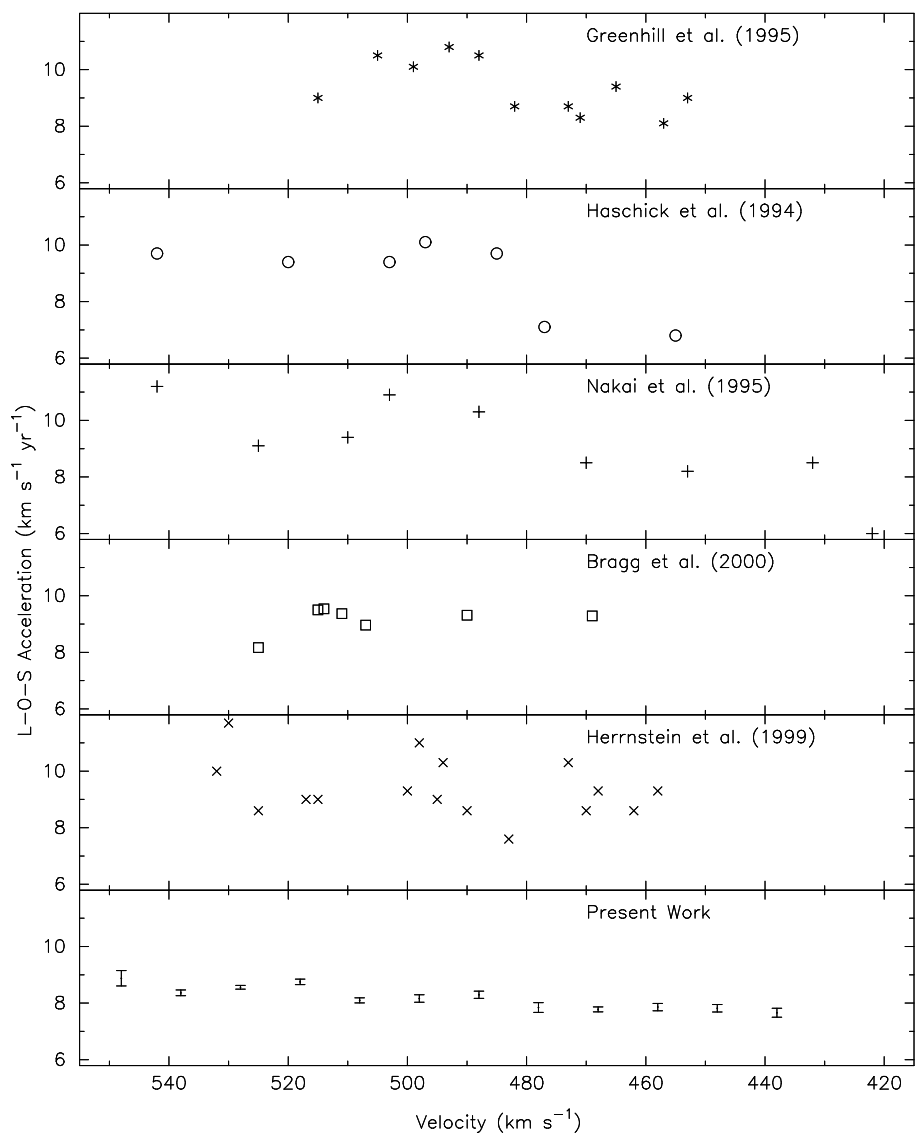}
\caption{Comparison of acceleration measurements for low-velocity maser 
spectral components. All studies are plotted
at the mid-date of their monitoring period, in order of increasing
mid-date. Only components for which sufficient velocity drift data
were obtained have been plotted here. The
majority of studies show a trend of increasing acceleration
as a function of Doppler velocity.
We note that the present work is represented by binned data here, whereas
other studies display measurements for individual components. The uncertainties
plotted for the binned values are the standard deviation of the mean of each
bin, as for Figure~\ref{f:systrend}b. For comparison, in the present work we
measure accelerations for $\sim$55 components in the low-velocity spectrum.
\label{f:lowvelacccomp}}
\end{figure}

\clearpage

\begin{figure}
%\includegraphics[angle=-90,scale=0.2]{newherrndisk.eps}
%\plotone{newherrndisk.eps}
\plotone{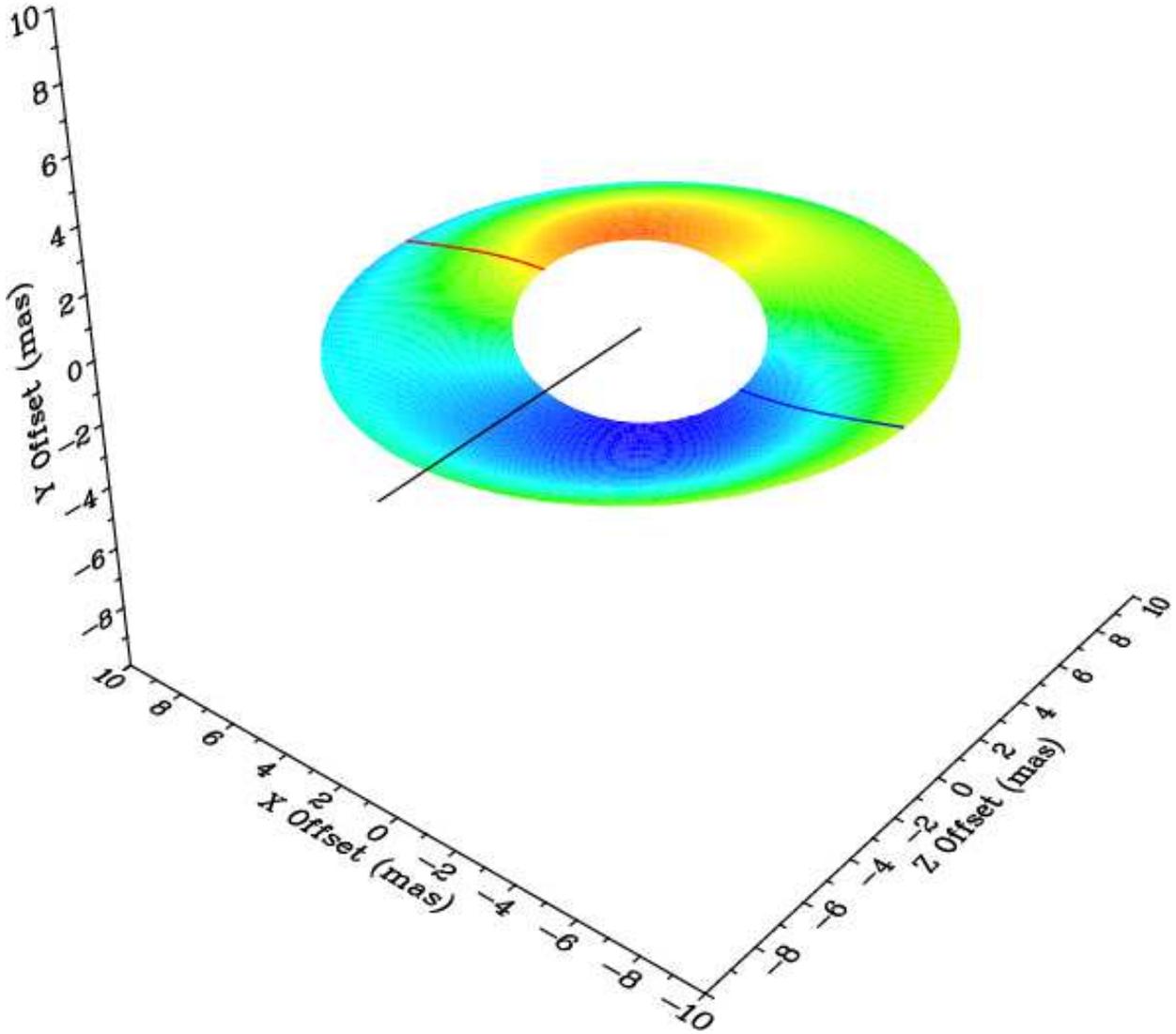}
\caption{Best-fitting maser disk from \citet{Herrnstein2005} viewed from [-40,40,-50]. 
The line-of-sight is shown as a line extending beyond the outer edge of the disk in black along the $z$-direction.  
Solid color contours show disk elevation, where red is the maxmium and
dark blue is the minimum. The disk midline for $\phi$=0\degrs and  180\degrs
is shown for red-shifted emission (red line) and blue-shifted emission
(blue line) respectively. Low-velocity (systemic) masers lie in the concavity on the front
side of the disk. 
\label{f:herrndisk}}
\end{figure}

\clearpage

\begin{figure}
\vspace{-4cm}
\epsscale{0.7}
%\plotone{jrh_and_our_masers.ps}
\plotone{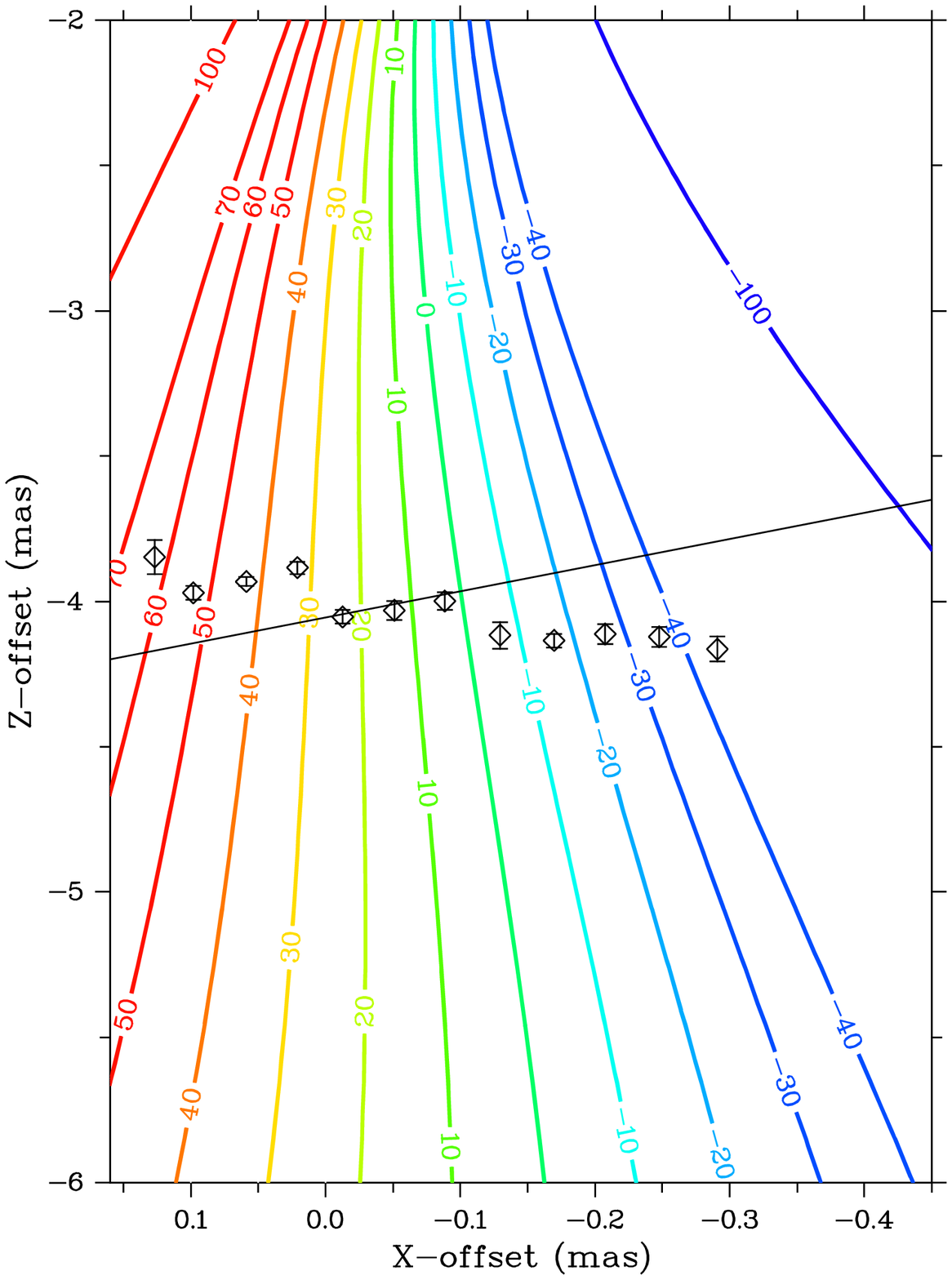}
\caption{Location of the low-velocity emission in the concavity \citep{Herrnstein2005}.
Isovelocity contours are marked in color and are in units of \velunits relative
to v$_{sys}$ \velunits \citep{Herrnstein2005}. The LOS is along the $z$-direction, and
therefore a vertical contour would mark a zero velocity gradient in the line-of-sight. The black
line marks the bottom of the concavity. The black symbols mark the position of the low
velocity emission, derived from the acceleration data. Observer views along the $z$-axis.
\label{f:velcontours}}
\end{figure}

\clearpage

\begin{figure}
\includegraphics[angle=-90,scale=0.8]{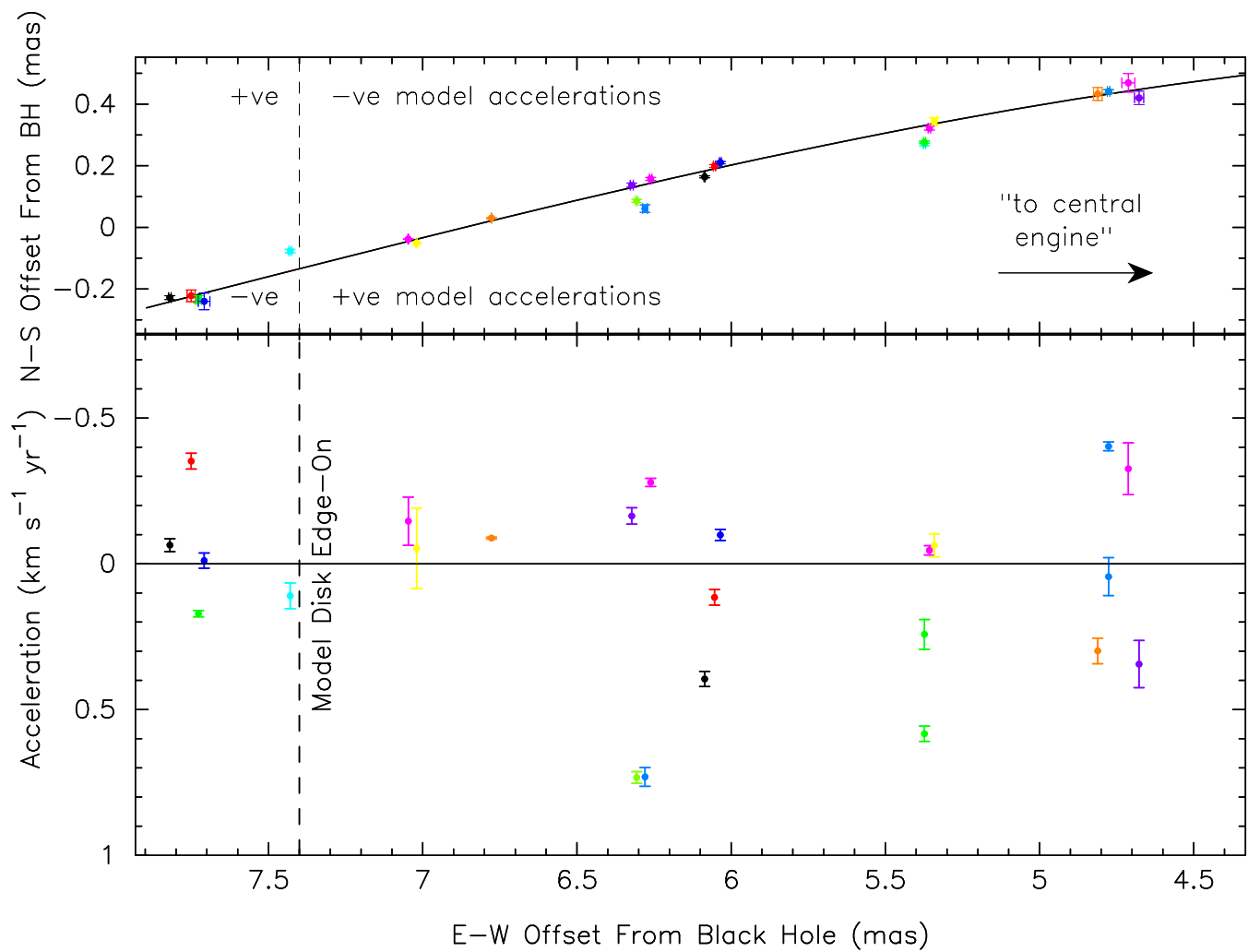}
\caption{Sky positions and accelerations of fitted red-shifted maser components.
{\it (Lower Panel)} LOS accelerations plotted as a function of
disk east-west offset. The solid line marks zero LOS acceleration,
which should occur when masers are on the disk midline.
{\it (Upper Panel)} Sky positions of the corresponding maser components.
The solid line indicates the disk midline predicted by the
geometric model of \citet{Herrnstein2005}. All maser positions
Note that, for this model, masers above the model midline {\it (upper panel)} 
should also have negative accelerations at radii $<$ 7.4~mas, and positive 
accelerations at greater radii.
\label{f:redsky}}
\end{figure}

\clearpage

\begin{figure}
%\epsscale{.80}
%\includegraphics[angle=-90,scale=0.8]{bluedisk.cps}
\includegraphics[angle=-90,scale=0.8]{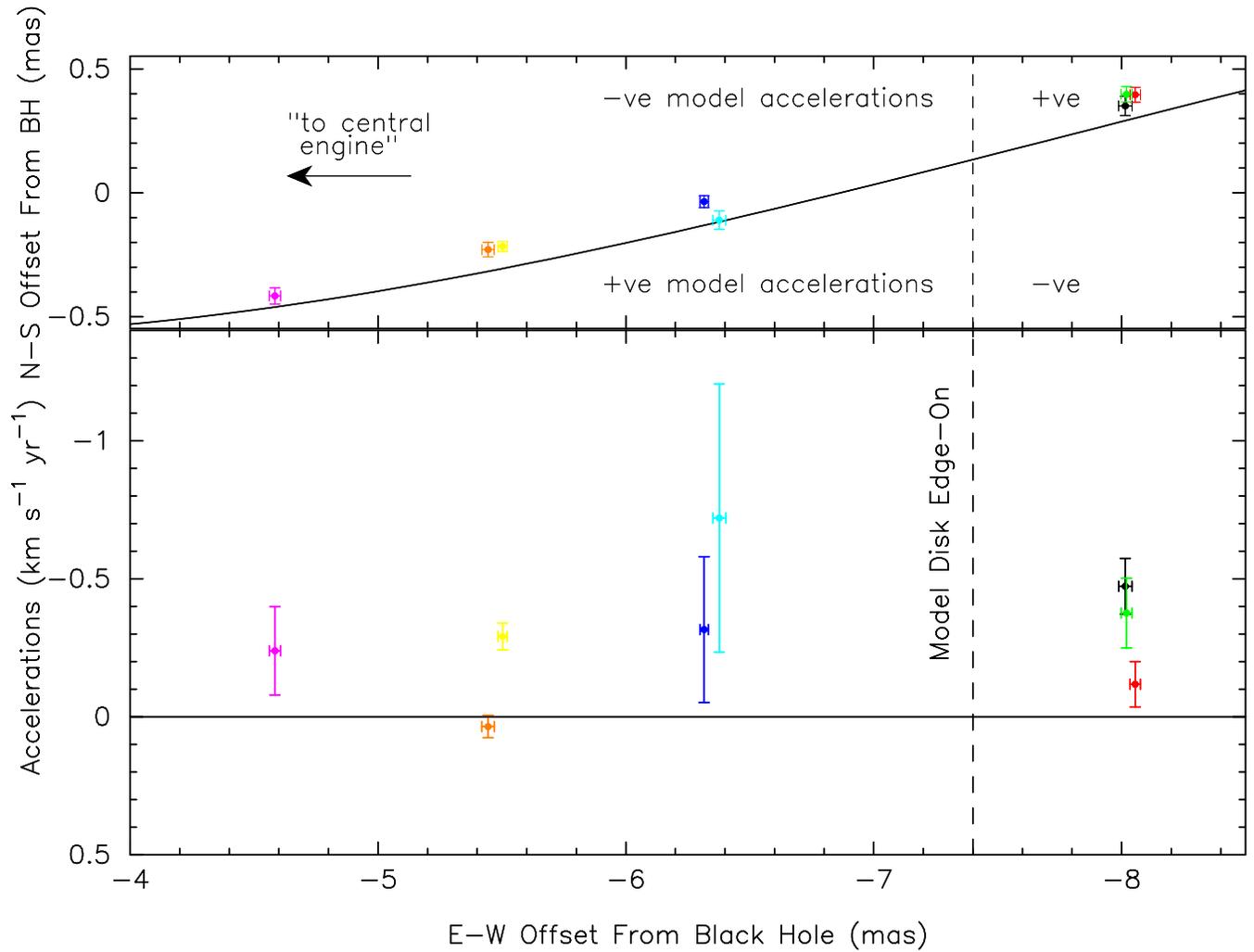}
\caption{As for Figure~\ref{f:redsky}, but for fitted blue-shifted high-velocity 
components. For this model, masers above the model midline {\it (upper panel)}  should have 
negative accelerations at radii $<$ 7.4~mas, and positive accelerations at greater radii.
\label{f:bluesky}}
\end{figure}

\clearpage

\begin{figure}
\epsscale{.80}
%\plotone{hv_all.ljgedt.eps}
\plotone{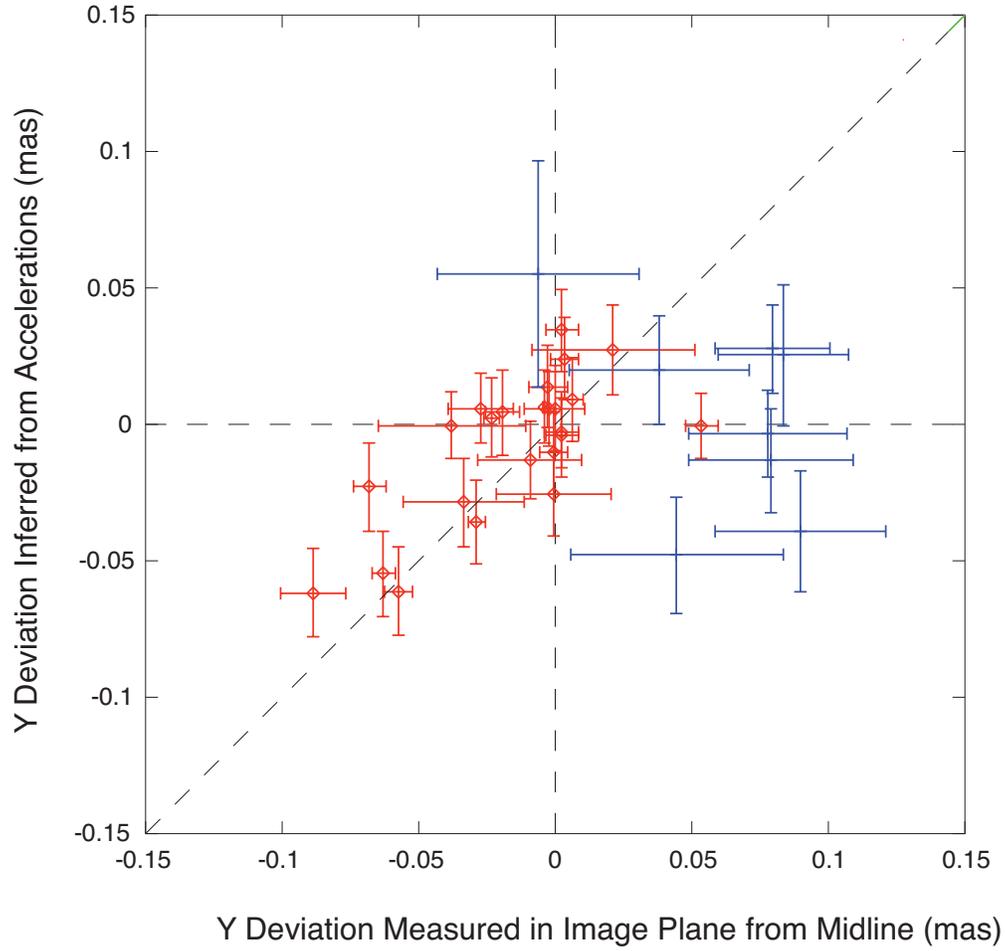}
\caption{Vertical deviations from the warped disk of \citet{Herrnstein2005}
calculated using acceleration data and VLBI data. See Section~6.1 for details.
\label{f:jim}}
\end{figure}
\clearpage

\begin{figure}
\hspace{3cm}
\includegraphics[angle=-90,scale=0.8]{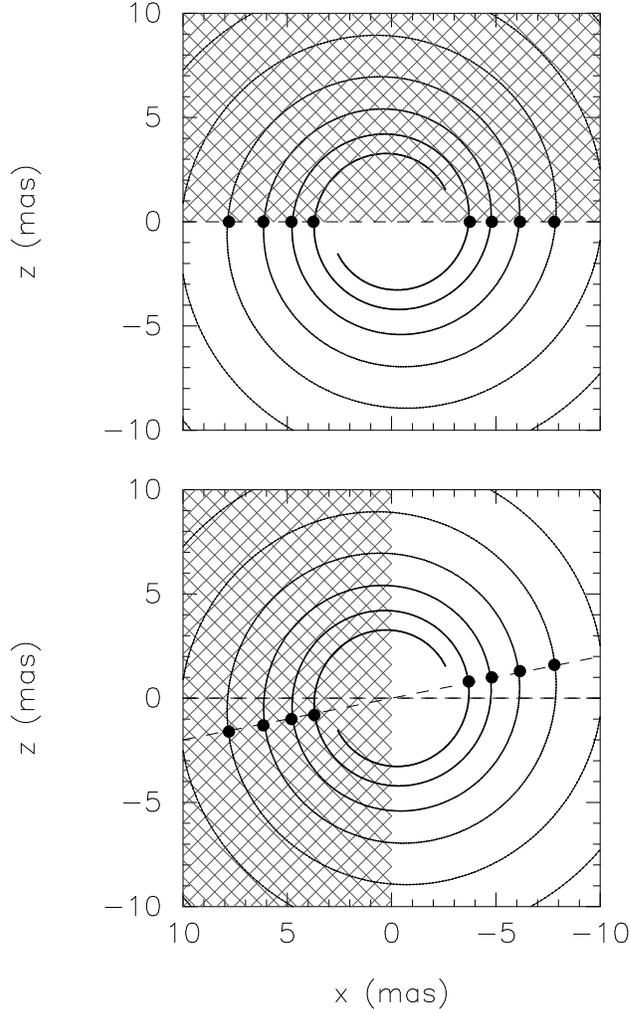}
\caption{
Comparison of spiral structure model predictions in the 
NGC~4258 accretion disk. {\it (Upper)} Schematic of maser disk according 
\citet{Maoz1995}. Masers (filled circles) are periodically located in density peaks 
 of trailing spiral structures (represented here by a 2-arm 
logarithmic spiral). 
Accelerations are negative behind the midline (hatched area) and 
positive in front of the midline. Components should occur within 
several degrees of the midline for maximum velocity coherent maser path lengths. 
{\it (Lower)} Schematic of the maser disk according to MM98. Masers (filled circles) 
are periodically located in the disk, but offset from the midline by 2\degrs (note the 
angle is exaggerated here). Red-shifted high-velocity components have negative accelerations 
(hatched area), whereas blue-shifted component accelerations should be positive 
statistically-speaking. Maser emission occurs at the longest velocity coherent 
path lengths along the line-of-sight through spiral arms.
\label{f:spiralcomp}}
\end{figure}

\clearpage

\begin{figure}
\includegraphics[angle=-90,scale=0.7]{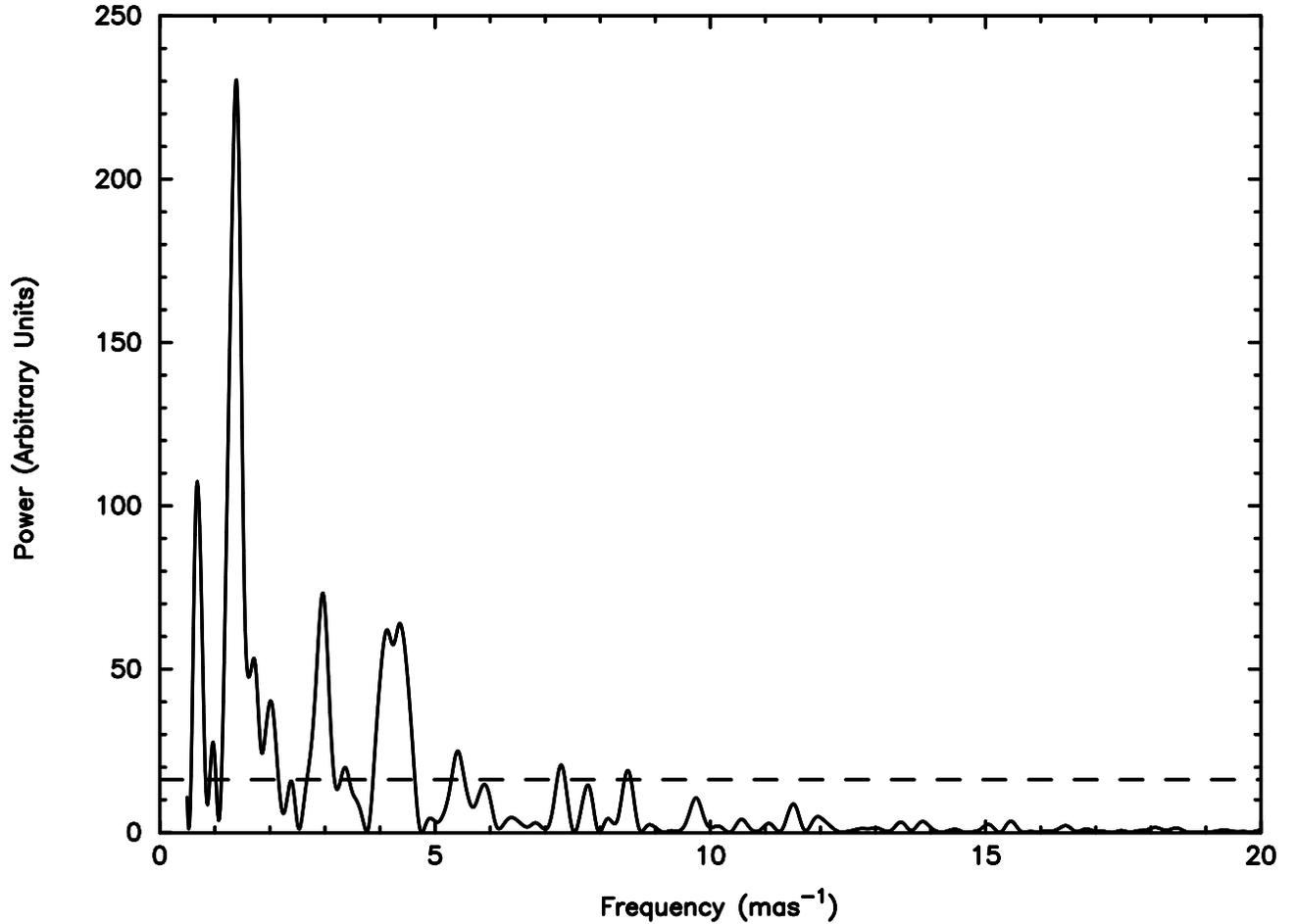}
\caption{
Periodogram of the positions of the high-velocity components in the NGC~4258 disk. The dominant frequency peak at 1.33 mas$^{-1}$ corresponds to a characteristic wavelength of 0.75 mas. The dashed horizontal black line marks the 3$\sigma$ significance level. We confined our analysis to ``wavelengths'' of  $<$ 2 mas, since longer periods would not be well-constrained by our dataset.
\label{f:periodogram}}
\end{figure}

\clearpage

\begin{figure}
\hspace{0.7cm}
\includegraphics[angle=-90,scale=.90]{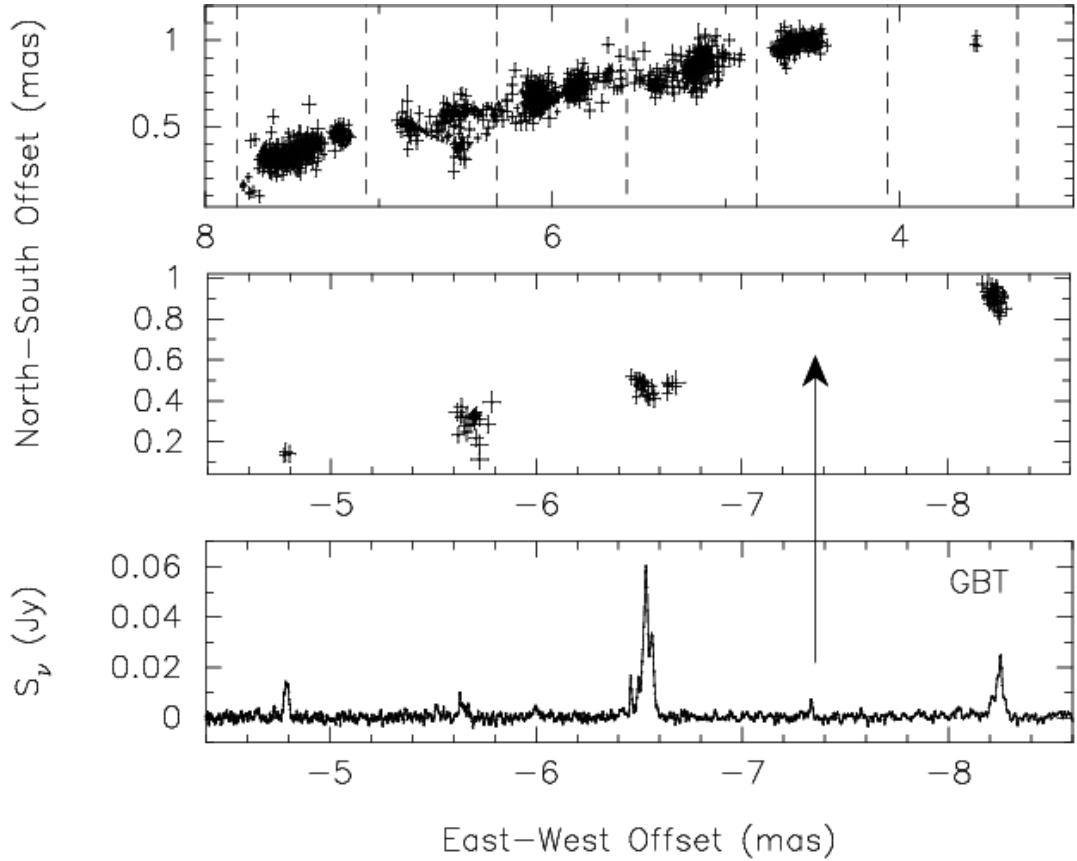}
\caption{{\it (Top)} Sky positions of red-shifted high-velocity 
emission in the NGC~4258 disk obtained using VLBI over 18 epochs. 
Dashed lines mark nodes at characteristic wavelengths of 0.75 mas, 
proposed by \citet{Maoz1995}. {\it (Middle)} Blue-shifted components in the 
NGC~4258 disk. {\it (Bottom)} GBT spectrum of the blue-shifted high-velocity 
emission taken on 2003 October 23. Velocity has been represented as approximate 
east-west offset using offset . The arrow marks the velocity of the ``missing'' blue 
component predicted by the Fourier analysis and detected independently in the spectrum.
\label{f:periodicityprediction}
}
\end{figure}

\clearpage

\begin{figure}
\hspace{0.7cm}
%\vspace{-1.0cm}
%\includegraphics[angle=-90,scale=.75]{nbody_4masses_3epochs.ps}
\includegraphics[angle=-90,scale=.75]{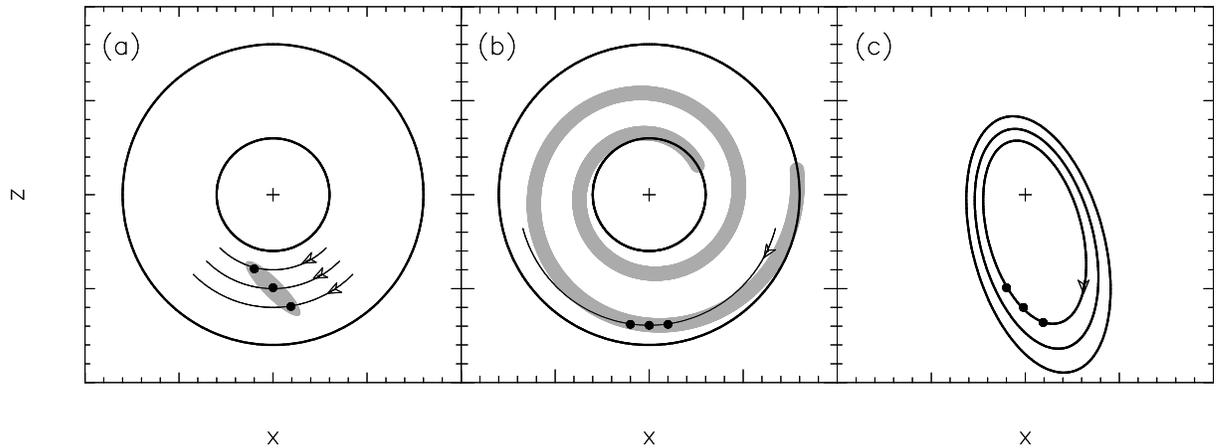}
\caption{{\footnotesize
Scenarios for producing the trend in line-of-sight accelerations
of low-velocity emission as a function
of component Doppler velocity. The line-of-sight direction is along the $z$-axis.
{\it (a)} Masers in circular, Keplerian orbits with locations determined by disk geometry and orientation. 
In this scenario, components (shown as filled, black circles) are in circular
orbits about the black hole (black cross) but only produce visible maser action 
within a favorable locus of disk radius and azimuth angle (shown in gray) determined by
the warping of the disk and tangency of the line of sight to the disk.  The
acceleration trend in this case is due to emission from a number of different
maser components in orbits of different radii.
{\it (b)} Masers in a circular, Keplerian orbit that crosses a spiral arm.
In this scenario, maser components are in a circular
orbit about the black hole and encounter a spiral
arm that perturbs the orbital motions (the arm is represented here by a gray logarithmic spiral, but 
a more realistic scenario would be fragmentary, as in flocculent spiral galaxy disks.) 
The acceleration trend in this case is due
to the passage of maser components through a spiral arm of lower
pattern speed than the local Keplerian velocity (see Figure~\ref{f:spiralacc}). 
{\it (c)} Masers in non-circular, confocal, Keplerian orbits. Maser components in the
same elliptical orbit by definition are at different radii and exhibit different 
accelerations, but they may instead lie on different orbits, with the same effect.}
\label{f:threepanel}}
\end{figure}

\clearpage

\begin{figure}
\hspace{0.7cm}
%\vspace{-1.0cm}
%\includegraphics[angle=-90,scale=.75]{nbody_4masses_3epochs.ps}
\includegraphics[angle=-90,scale=.75]{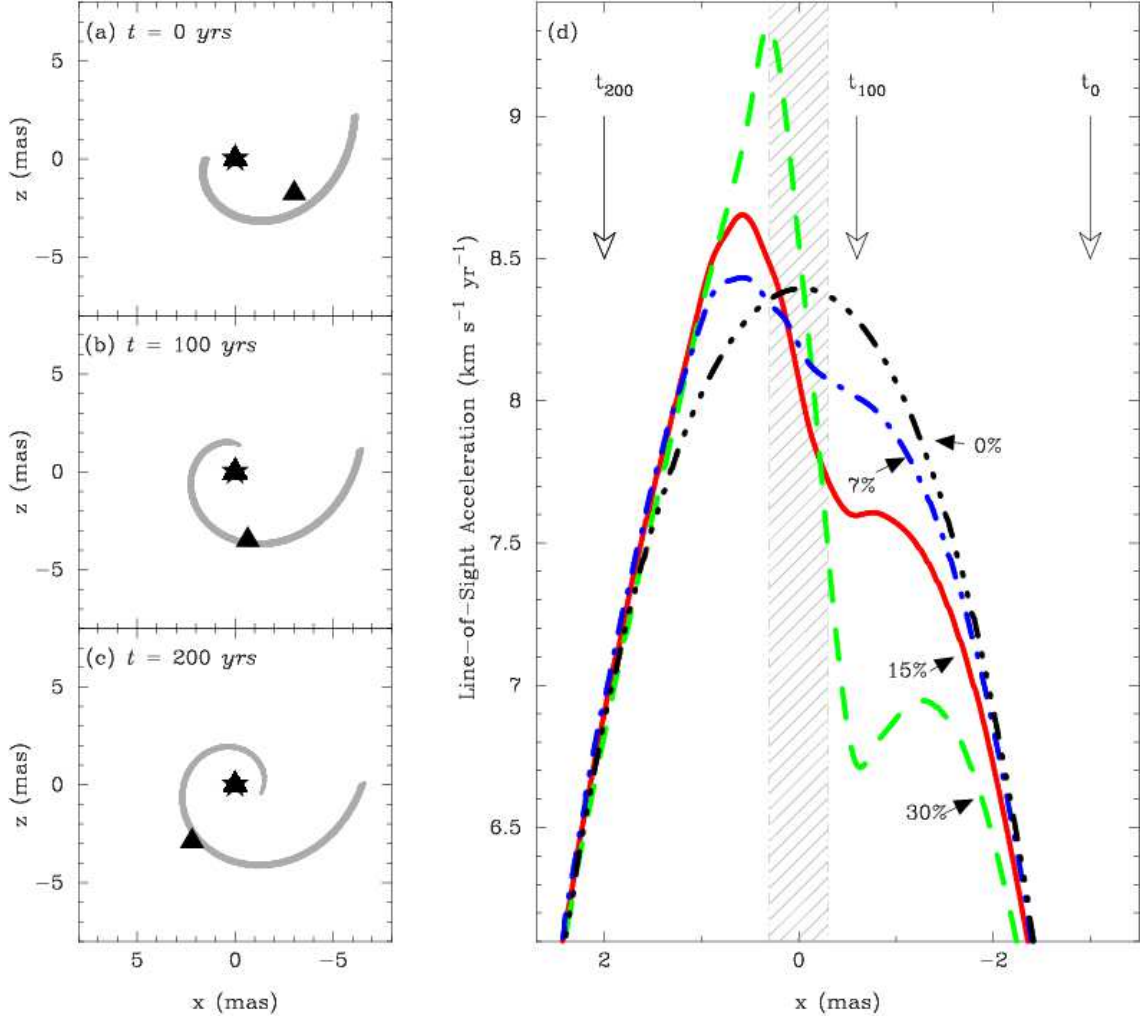}
\caption{{\footnotesize
The effect of a spiral arm on accelerations of low-velocity
masers in the NGC~4258 disk.{\it (a)} Initial conditions for an N-body simulation
of a maser clump (black triangle) orbiting a supermassive black hole of $3.8 \times 10^7$ $M_{\odot}$
\citep{Herrnstein2005} (black star). The parameters of the trailing logarithmic spiral arm (grey),
represented by 8000 particles, were arbitrarily chosen to be r$_{0}$ = 1.5 mas, $b$=0.42 and
$\phi_{0}$ = 0\degrs, where  $r = r_0 \exp (b (\phi-\phi_{0}))$. The pattern speed of the spiral arm
is half that of the Keplerian orbital velocity at any given radius.
Note that the true breadth of the arm (10$^{-2}$ pc) is exaggerated for clarity.
{\it (b)} Snapshot at t = 100 years.
{\it (c)} Snapshot at t = 200 years.
{\it (d)} Maser component line-of-sight acceleration as a function
of disk impact parameter due to both the supermassive black hole and a spiral arm of
varying masses. Arm mass is quoted as a percentage of the upper limit maser disk mass,
M$_{disk,upper}$ = 9 $\times$ 10$^5$ M$_{\odot}$ \citep{Herrnstein2005}. Note the significant deviation of the acceleration
profiles at 7, 15 and 30\% from that of the 0\% (black hole only) sinusoid. The curve that best matches
observations is that for an arm mass of 15\%. The grey hatched area indicates the observed projected extent in disk
impact parameter of the low velocity masers.}
\label{f:spiralacc}}
\end{figure}

\clearpage

\begin{figure}
\includegraphics[angle=-90,scale=.70]{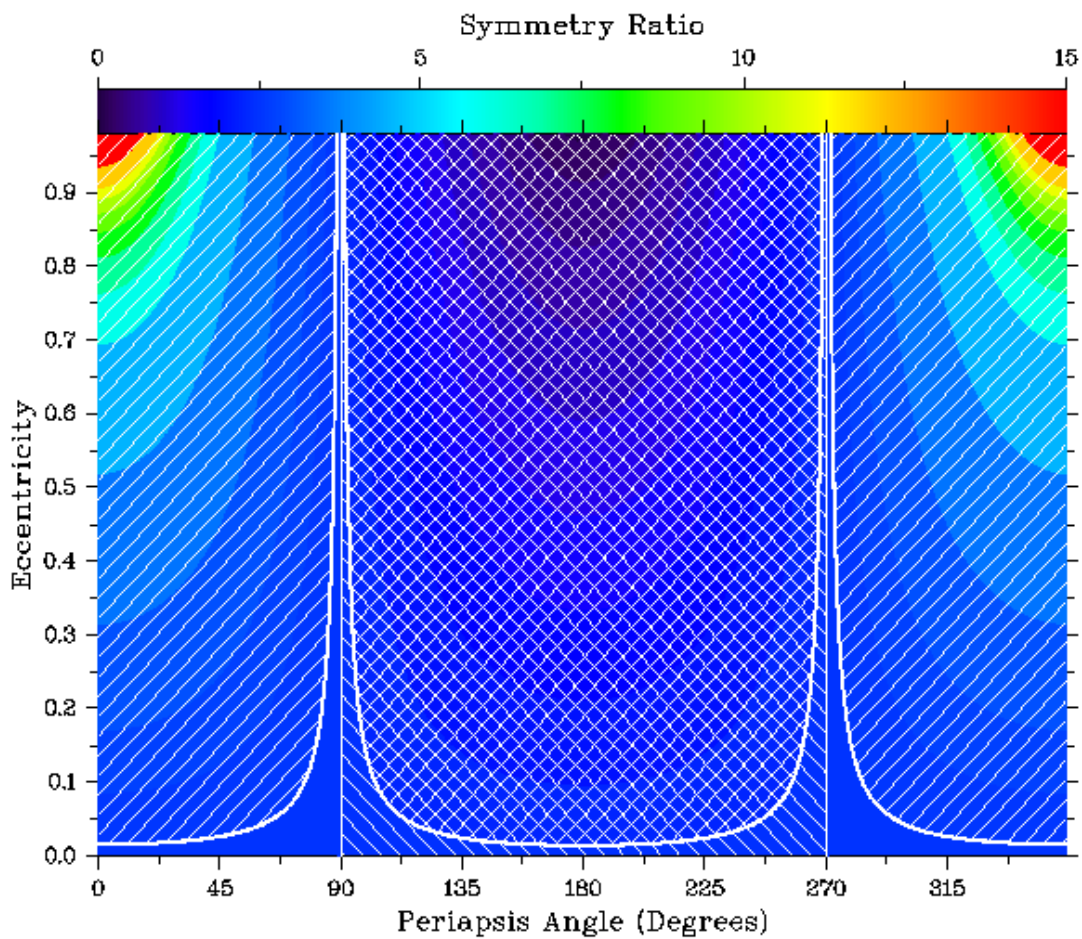}
\caption{Contour plot of the ``symmetry'' ratio ($v_{red,\,los}/v_{blue,\,los}$) of model P-V diagrams for
eccentricity ($e$) as a function of periapsis angle ($\omega$). The symmetry of the data 
eliminates values of $e$ and $\omega$ in the positive-gradient hatched regions of the plot.
Using acceleration data for low-velocity emission, we further eliminate all of the negative-gradient
hatched region of 90\degrs $\le$ $\omega$ $\le$ 270\degrs for low-velocity maser components 
on the same orbit. Only unhatched regions of parameter space are permitted by our dataset.
\label{f:stretch}}
\end{figure}

%% If you are not including electonic art with your submission, you may
%% mark up your captions using the \figcaption command. See the
%% No more than seven \figcaption commands are allowed per page,
%% so if you have more than seven captions, insert a \clearpage

%% Tables should be submitted one per page, so put a \clearpage before

%% Two options are available to the author for producing tables:  the
%% deluxetable environment provided by the AASTeX package or the LaTeX
%% table environment.  Use of deluxetable is preferred.
%%

%% Three table samples follow, two marked up in the deluxetable environment,
%% one marked up as a LaTeX table.

%% In this first example, note that the \tabletypesize{}
%% command has been used to reduce the font size of the table.
%% We also use the \rotate command to rotate the table to
%% landscape orientation since it is very wide even at the
%% Note also that the \label command needs to be placed

%% This table also includes a table comment indicating that the full
%% version will be available in machine-readable format in the electronic

\end{document}